\documentclass[aps,prd,preprint,floatfix,nofootinbib,superscriptaddress]{revtex4}
\usepackage{graphicx,subfigure,epsfig,epsf}
\newcommand{\beq}{\begin{equation}}
\newcommand{\eeq}{\end{equation}}
\newcommand{\bea}{\begin{eqnarray}}
\newcommand{\eea}{\end{eqnarray}}
\newcommand{\bec}{\begin{center}}
\newcommand{\eec}{\end{center}}
\newcommand{\lvm}{\leavevmode}

\def\etal{{\it et al.}}
\def\to{\rightarrow}

\def\psla{\big/\!\!\!p}
\def\ksla{\big/\!\!\!k}

\def\kprsla{\big/\!\!\!k^{'}}

\begin{document}

\def\issue(#1,#2,#3){{\bf #1,} #2 (#3)} 
\def\APP(#1,#2,#3){Acta Phys.\ Pol.\ B \ \issue(#1,#2,#3)}
\def\ARNPS(#1,#2,#3){Ann.\ Rev.\ Nucl.\ Part.\ Sci.\ \issue(#1,#2,#3)}
\def\CPC(#1,#2,#3){Comp.\ Phys.\ Comm.\ \issue(#1,#2,#3)}
\def\CIP(#1,#2,#3){Comput.\ Phys.\ \issue(#1,#2,#3)}
\def\EPJC(#1,#2,#3){Eur.\ Phys.\ J.\ C\ \issue(#1,#2,#3)}
\def\EPJD(#1,#2,#3){Eur.\ Phys.\ J. Direct\ C\ \issue(#1,#2,#3)}
\def\IEEETNS(#1,#2,#3){IEEE Trans.\ Nucl.\ Sci.\ \issue(#1,#2,#3)}
\def\IJMP(#1,#2,#3){Int.\ J.\ Mod.\ Phys. \issue(#1,#2,#3)}
\def\JHEP(#1,#2,#3){J.\ High Energy Physics \issue(#1,#2,#3)}
\def\JPG(#1,#2,#3){J.\ Phys.\ G \issue(#1,#2,#3)}
\def\MPL(#1,#2,#3){Mod.\ Phys.\ Lett.\ \issue(#1,#2,#3)}
\def\NP(#1,#2,#3){Nucl.\ Phys.\ \issue(#1,#2,#3)}
\def\NIM(#1,#2,#3){Nucl.\ Instrum.\ Meth.\ \issue(#1,#2,#3)}
\def\PL(#1,#2,#3){Phys.\ Lett.\ \issue(#1,#2,#3)}
\def\PRD(#1,#2,#3){Phys.\ Rev.\ D \issue(#1,#2,#3)}
\def\PRL(#1,#2,#3){Phys.\ Rev.\ Lett.\ \issue(#1,#2,#3)}
\def\PTP(#1,#2,#3){Progs.\ Theo.\ Phys. \ \issue(#1,#2,#3)}
\def\RMP(#1,#2,#3){Rev.\ Mod.\ Phys.\ \issue(#1,#2,#3)}
\def\SJNP(#1,#2,#3){Sov.\ J. Nucl.\ Phys.\ \issue(#1,#2,#3)}
\bibliographystyle{revtex}

\title{TeV Scale Implications of  Non Commutative Space time  in Laboratory Frame with Polarized Beams}

\author{Sumit K. Garg}
\email{sumit@cts.iisc.ernet.in}
\affiliation{Centre for High Energy Physics, Indian Institute of Science, Bangalore 560 012, India}

\author{T. Shreecharan}
\email{shreet@imsc.res.in}
\affiliation{School of Physics, University of Hyderabad, Hyderabad 500
046, India }

\author{P. K. Das}
\email{pdas@bits-goa.ac.in}
\affiliation{Department of Physics, Birla Institute of Technology and Science-Pilani, K.K. Birla Goa campus, NH-17B, Zuarinagar, Goa 403 726, India}

\author{N. G. Deshpande}
\email{desh@uoregon.edu}
\affiliation{Institute of Theoretical Science, University of Oregon, Eugene, Oregon 97403, USA}

\author{G. Rajasekaran}
\email{graj@imsc.res.in}
\affiliation{The Institute of Mathematical Sciences, C.I.T. Campus, Taramani, Chennai 600 113, India}
\affiliation{ Chennai Mathematical Institute, Siruseri 603103, India}

\begin{abstract}
We analyze $e^{+}e^{-}\rightarrow \gamma\gamma$, 
$e^{-}\gamma \rightarrow e^{-}\gamma$ and $\gamma\gamma \rightarrow e^{+}e^{-} $  processes  within the Seiberg-Witten 
expanded  noncommutative  scenario using polarized beams. 
With unpolarized beams the leading order effects of non commutativity  
starts from  second order in non commutative(NC) parameter i.e. $O(\Theta^2)$, while with polarized beams  
these corrections appear at first order  ($O(\Theta)$) in cross section. The corrections in Compton case  can probe the magnetic component($\vec{\Theta}_B$) 
while in Pair production and Pair annihilation  probe the electric component($\vec{\Theta}_E$) of NC parameter. We include the effects of earth
rotation in our analysis. This study is done by investigating the effects of non commutativity on different time averaged cross section observables. The  results which also depends on the position
of the collider,  can provide clear and distinct signatures of the model testable at the
 International Linear Collider(ILC).


\end{abstract}
\pacs{{11.10.Nx.}}

\maketitle

\section{Introduction}
Field theories defined on a non commutative(NC) space time have been extensively 
studied in the past few years. This idea received much attention due to its possible 
connection with quantum gravity and because of its natural origin in
 string theories. Infact Seiberg and Witten\cite{bergwitten} described how
NC gauge theory can emerge as a low energy manifestation of string theory.

However the original idea was considered long time ago when non commutativity of Minkowski Space-time was
assumed as a natural extension of Heisenberg's position-momentum non commutativity in quantum mechanics. 
In early work of Snyder\cite{snyderUV} the non commutativity of space time  was  
suggested as a possible cure for ultraviolet divergences in Quantum 
field theory(QFT). However this viewpoint was largely  ignored mainly because  at that time  
the renormalization
techniques in QFT met great success predicting quite accurately the numerical values
for physical observables in quantum electrodynamics(QED).  

This field got renewed attention  in 2000 after the work of Seiberg and Witten\cite{bergwitten}. They showed that the dynamics
of open strings ending on D-branes in a background field can be described
by a non commutative quantum field theory(NCQFT). They also gave the explicit mapping between NC and ordinary
gauge theories which is famously known as the Seiberg-Witten Map(SWM). This mapping in turn emerged as the 
roadmap for investigation of 
the gauge theories like Standard Model in non commutative space time.

Parallel to this development another approach, based on the Moyal-Weyl (MW) star product Eq.(\ref{StarP}), also became popular. It was very soon realized that NC field theories constructed via this approach are plagued by the so called UV/IR effect \cite{Minwalla,Susskind} wherein additional divergences appeared in the infrared that were not present in the commutative theory.

Apart from these developments NC theories are also
supposed to  shed some light 
on the quantization of space time (i.e. quantum theory of gravity) in the context of string theory. Thus it served as one of the major motivation for the intense activity in this area among string theorists. These ideas may lead to  
 the possibility of making the Standard Model(SM) consistent with quantum gravity. In this context it worth pointing out that that the UV/IR effect plays an important role in determining what the UV theory might be. In other words a high energy theory might show up consequences at low energy which are will within experimental reach. This has been discussed, within the context of MW NC setting, in the works \cite{Alvarez,Jaeckel1,Jaeckel2,Horvat}.

Hence keeping in mind the above motivations it is reasonable to examine field theories,
 and in particular 
the standard model of particle physics on non commutative space time. We adopt an
approach based on SWM popularized by the Munich group\cite{Wess1,Wess2,Calmetetal, NGetal, Duplancicetal, ggotheta, Melicetal1,Melicetal2}.


The reason why NC collider phenomenology is interesting, comes from the fact that the scale of non commutativity could be as low as a few TeV, which can be explored at the present or the future colliders.
This led to a great deal of interest in phenomenology  of the NCSM with SWM. Many phenomenological signatures  have been studied by different research groups. These works were mainly
done \cite{Dasetal,Hewett01,Schuppgneutr,Haghighatgneutr,Najafabadipoltopqur,MahajantbW,IltanZdecays,Deshpandetriplecoup,
Najafabaditopqur,Trampeticpheno,BuricZgg,Abbiendipairann,Melicquarkonia,Melickpi,Alboteanuothsq,Prasanta1,Prasanta2,wang} with unpolarized beams with leading corrections to SM starting from $O(\Theta^2)$. However few studies \cite{ggotheta, AlboteanuLHC,AlboteanuILC} are also done  with corrections at the $O(\Theta)$ in cross section. Previous studies for processes considered here are often incomplete because $O(\Theta^2)$ contribution to scattering amplitudes requires Feynman rules to $O(\Theta^2)$ which were not included, and these terms are known to have intrinsic ambiguities and thus making the calculations indefinite.

In this work we have calculated ($O(\Theta)$) corrections for Compton, pair annihilation and pair production while 
keeping only one initial beam polarization($e^{-}$ in Compton and pair annihilation and $\gamma$ in pair production). 
We have also taken into account the effect of earth's rotation\cite{earthroteffs1, earthroteffs2, earthroteffs3, earthroteffs4} on 
observable signals of NC. The effects of Non commutativity is studied on various time averaged observables to determine  the magnitude 
and direction of NC parameter.

We have looked at the possible implications of the NC corrections for phenomenology at the International Linear
Collider(ILC)\cite{ILC1,ILC2}. In addition to $e^{-}e^{+}$ programme, linear colliders
also provide a unique opportunity to study $\gamma \gamma$ and $\gamma e$ interactions at energies and luminosities 
comparable  to initial electron-positron beam. Intense beam of high energy photons can be obtained using Compton backscattering
of laser light off the high energy electrons. If these beams become available at future Linear Colliders then it  
will serve as crucial test
of NCQED in the processes we discuss.

The rest of the paper is organized as follows. In section II, we will briefly describe the mathematical description of non 
commutative space time.
In section III, we give the cross section details for the mentioned processes. In section IV, 
we will present our numerical results. Finally we conclude with a section on our results and a discussion.

\section{Non Commutative Standard Model}
The idea of non commutative space is the generalization of quantum mechanics
in which the canonical position and momentum variables {$x_i, p_j$} are replaced with hermitian
operators {$\hat{x}_i, \hat{p}_j$} which obey the famous Heisenberg commutation relation
\[  [\hat{x}_i, \hat{p}_j] = \frac{i h}{(2 \pi)} \delta_{ij}\] 

So just like the qunatization of Classical Phase space, a non commutative space time coordinates $x^{\mu}$ are replaced by
the Hermitian generators $\hat{x}^{\mu}$ which obeys the NC commutation relations($\mu,\nu=0,1,2,3$)
\beq \label{NCST}
[\hat{x}^\mu,\hat{x}^\nu]=i\Theta^{\mu\nu}  
\eeq
where $\Theta^{\mu \nu}$ is antisymmetric constant matrix with units of (Length$)^2$. 
Thus in general one can consider two cases: first with space-space non commutativity associated 
with $\Theta^{ij}$, i,j=1,2,3 (known as magnetic  components) and second with space-time non 
commutative related with $\Theta^{0i}$ (electric components).

In field theory context, we realize Eq.(\ref{NCST}) by using Moyal-Weyl(MW) $\star$-product, defined by
\begin{equation} \label{StarP}
(f\star g)(x)= \exp\left(\frac{i}{2}\Theta_{\mu\nu}\partial^\mu_x\partial^\nu_y \right)f(x)g(y)|_{y=x}.
\end{equation}
Thus one can construct NC field theories by replacing the ordinary products of fields with the corresponding star products. This replacement affects only the interaction parts and not the free field theory. In case of gauge theories, this approach is only consistent for $U(N)$ gauge theories and only a single eigenvalue is allowed for the charge operator.


To construct the NC extension of the standard model (SM) \cite{Calmetetal,NGetal,Melicetal1,Melicetal2}, which uses the same gauge group and particle content, or for that matter any other gauge theory, be it abelian or non-abelian, one expands the NC gauge fields in non linear power series of $\Theta$ \cite{bergwitten,Wess1,Wess2}:
\bea \label{swps}
\lambda_\alpha (x,\Theta) & = & \alpha(x) + \Theta^{\mu\nu} \lambda^{(1)}_{\mu\nu}(x;\alpha) + \Theta^{\mu\nu} \Theta^{\eta\sigma} \lambda^{(2)}_{\mu\nu\eta\sigma}(x;\alpha) + \cdots \\
A_\rho (x,\Theta) &=& A_\rho(x) + \Theta^{\mu\nu} A^{(1)}_{\mu\nu\rho}(x) + \Theta^{\mu\nu} \Theta^{\eta\sigma} A^{(2)}_{\mu\nu\eta\sigma\rho}(x) + \cdots .
\eea

At face value it can be seen from the above map that SW approach leads to a field theory with an infinite number of vertices and Feynman graphs thereby leading to an uncontrolled degree of divergence inturn giving an impression of complete failure of perturbative renormalization. But over the years a number of studies have shown that it is possible to construct anomaly free, renormalizable, and effective theories at one loop and first order in $\Theta$ \cite{Bichl1,Martin,Martin-Tam1,Buric1,Buric2,Buric3,Martin-Tam2,Ettefaghi,Buric4}. Before we provide the Feynman rules it must be mentioned that the celebrated IR/UV mixing, discussed in the earlier section, does not exist in the above $\Theta$ expanded approach. Though this is not a drawback in the scales of our interest there do exist certain phenomena that require all orders of the NC parameter be retained. This led to the so called $\Theta$-exact approach, that is from the exact solutions of the SW equations. The phenomenological consequences of this have been explored in \cite{Horvat2,Horvat3}.

\begin{figure}[t]
\begin{center}\lvm
\begin{minipage}[c]{160mm}
\begin{tabular}{cccc}
\begin{minipage}[c]{70mm}
\centerline{
\includegraphics[width=40mm, height=40mm, angle=0]{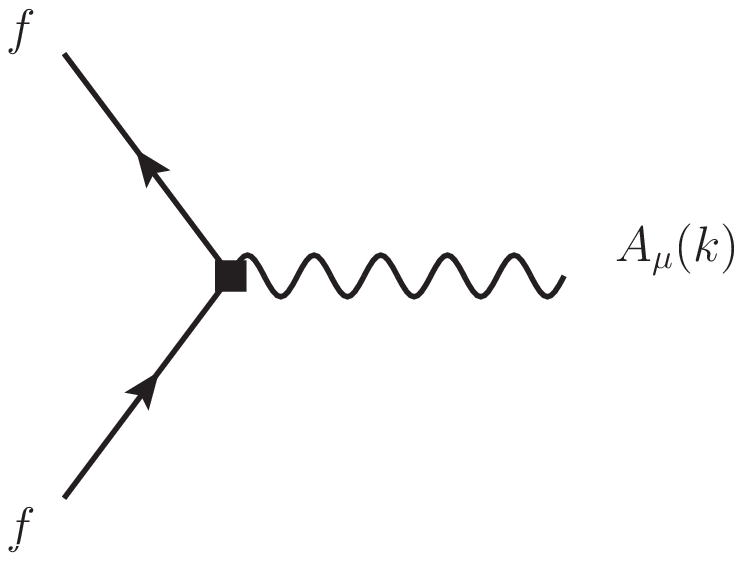}
}\vspace{-0.5cm}
\caption[]{3-point NC $e^{-}e^{-}\gamma $ vertex
             }
\label{fig:coordinates}
\end{minipage}
\begin{minipage}[c]{5mm}
\centerline{$ i e \Gamma_{\mu}$
}\vspace{-0.5cm}
\end{minipage}
\begin{minipage}[c]{70mm}
\centerline{
\includegraphics[width=50mm, height=40mm, angle=0]{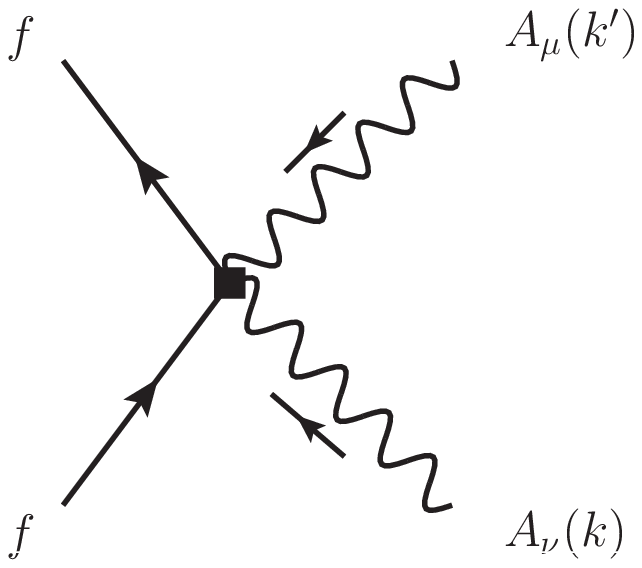}
}\vspace{-0.5cm}
\caption[]{4-point NC $e^{-}e^{-}\gamma \gamma $ vertex,
$Q_f$ denotes the electric charge of fermion f. }
\end{minipage}
&\quad
\begin{minipage}[c]{10mm}
\centerline{
$\frac{-e^2 Q_f^2}{2} \Theta_{\mu\nu\rho}(k^{'\rho} -k^{\rho})$
}
\end{minipage}
\end{tabular}
\end{minipage}\label{FeynCompton}
\end{center}
\end{figure}

The above mentioned studies provide confidence in using the using NC SW expanded SM for phenomenological purposes. The Feynman rules for the NCSM have been worked out in \cite{Calmetetal,Melicetal1,Melicetal2} and the ones relevant for this work specifically minimal NCSM are given below. The rules for the 3-point $e^{-}e^{-}\gamma $ vertex (Fig.1) and the 4-point $e^{-}e^{-}\gamma \gamma $ vertex (Fig.2) are \cite{Melicetal1}
\bea
{\Gamma}^{\mu} = \gamma^{\mu} -\frac{i}{2}\{ (p_{out}\Theta p_{in})\gamma^{\mu}-(p_{out}\Theta)^{\mu}(\psla_{in}-m)-(\psla_{out}-m)(\Theta p_{in})^{\mu}\}
\eea
and
\bea
\Theta^{\mu\nu\rho}(k^{'}-k)_{\rho}&=& \Theta^{\mu\nu}\gamma^{\rho}(k^{'}-k)_{\rho}+ \Theta^{\nu\rho}\gamma^{\mu}(k^{'}-k)_{\rho}+ \Theta^{\rho\mu}\gamma^{\nu}(k^{'}-k)_{\rho}\nonumber\\
&=& \Theta^{\mu\nu}(\kprsla-\ksla) + \gamma^{\mu}(\Theta (k^{'}-k))^{\nu} - (\Theta (k^{'}-k))^{\mu}\gamma^{\nu}
\eea
respectively. Here $p_{in}$ is the incoming and $p_{out}$ is outgoing momentum of fermion at vertex with
$(p_{out}\Theta p_{in})\equiv p_{\mu}\Theta^{\mu\nu}p_{\nu}$ , $(p\Theta)^{\nu} \equiv p_{\mu}\Theta^{\mu\nu}$ and $(\Theta p)^{\mu} \equiv \Theta^{\mu\nu}p_{\nu}$. However here we will work in the massless limit of electron and positron.

\section{Cross sections in the Laboratory Frame}
In this section we will give the calculational details of our work, first starting from the Compton 
scattering case. 





This process in NCQED proceed at the tree level by the following diagrams(Fig.3).
The first two diagrams also appears in pure QED while 3rd one arises just because of non commutative nature of space time
and is a contact interaction.

The Feynman amplitudes for these diagrams with initial $e^{-}$ beam polarization in 
NCQED are given by expressions:

\bea
i{\mathcal{M}}_{a} &=& \left[{\overline u}({p^{'}}) H_{L,R} (i e \Gamma^\mu) \frac{i}{\psla+\ksla} (i e \Gamma^\nu)u(p) \right]\epsilon_{r^{'}\mu}^{*}(k^{'})
 \epsilon_{r_\nu}(k) \nonumber \\
&=& \nonumber {\mathcal{M}}_{1a}\epsilon_{r^{'}}^{*}(k^{'})\epsilon_{r}(k) \nonumber\\
&=& \{ -e^2 T_{11}^{\mu \nu} + \frac{i e^2}{2} ((T_{12}^{\Theta})^{\mu\nu} + (T_{13}^{\Theta})^{\mu\nu})+ O(\Theta^2)\}\epsilon_{r^{'}\mu}^{*}(k^{'})\epsilon_{r_\nu}(k) \nonumber\\
\eea

\bea
i{\mathcal{M}}_{b} &=& \left[{\overline u}(p^{'}) H_{L,R} (i e \Gamma^\nu) \frac{i}{\psla-\kprsla} (i e \Gamma^\mu)u(p) \right]\epsilon_{r^{'}\mu}^{*}(k^{'})
 \epsilon_{r_\nu}(k) \nonumber \\
&=& \nonumber {\mathcal{M}}_{1b}\epsilon_{r^{'}}^{*}(k^{'})\epsilon_{r}(k) \nonumber\\
&=& \{ -e^2 T_{21}^{\nu \mu} + \frac{i e^2}{2} ((T_{22}^{\Theta})^{\nu\mu} + (T_{23}^{\Theta})^{\nu\mu})+ O(\Theta^2)\}\epsilon_{r^{'}\mu}^{*}(k^{'})\epsilon_{r_\nu}(k) \nonumber\\
\eea

\bea
i{\mathcal{M}}_{c} &=& \left[{\overline u}({p^{'}})\{ \frac{e^2}{2}\Theta^{\mu\nu\rho}(k^{'}+k)_{\rho} \} H_{(L,R)}  u(p) \right]
\epsilon_{r^{'}\mu}^{*}(k^{'}) \epsilon_{r_\nu}(k) \nonumber \\
&=& \nonumber {\mathcal{M}}_{1c}\epsilon_{r^{'}}^{*}(k^{'})\epsilon_{r}(k) \nonumber\\
&=& \{  \frac{i e^2}{2} (-i T^{\Theta})^{\mu\nu} \}\epsilon_{r^{'}\mu}^{*}(k^{'})\epsilon_{r_\nu}(k) \nonumber\\
\eea

\begin{figure}[t]
\begin{center}\lvm
\begin{minipage}[c]{160mm}
\begin{tabular}{ccc}
\begin{minipage}[c]{55mm}
\centerline{
\includegraphics[width=40mm, height=40mm, angle=0]{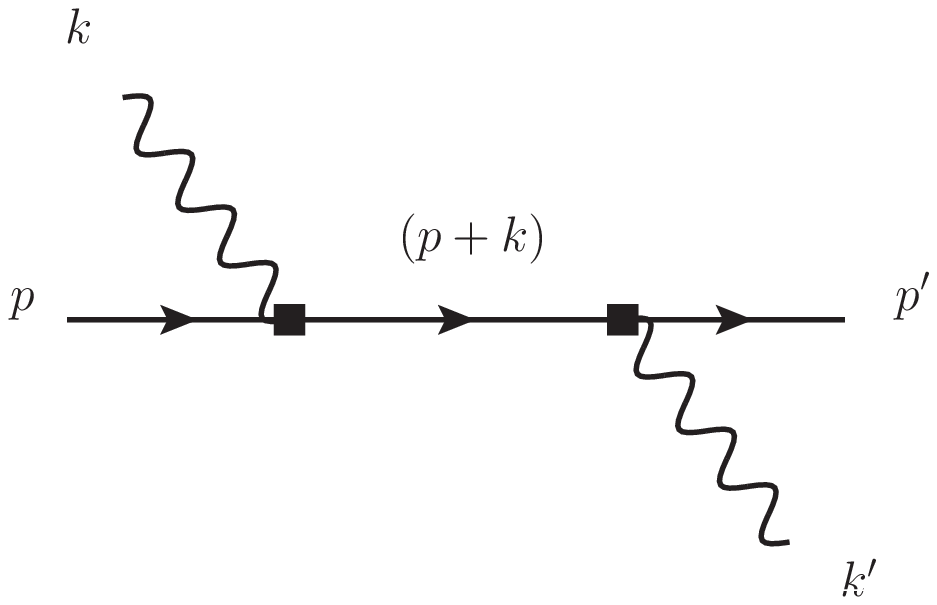}
}\vspace{-0.5cm}
\label{fig:coordinates}
\end{minipage}
\begin{minipage}[c]{55mm}
\centerline{
\includegraphics[width=40mm, height=40mm, angle=0]{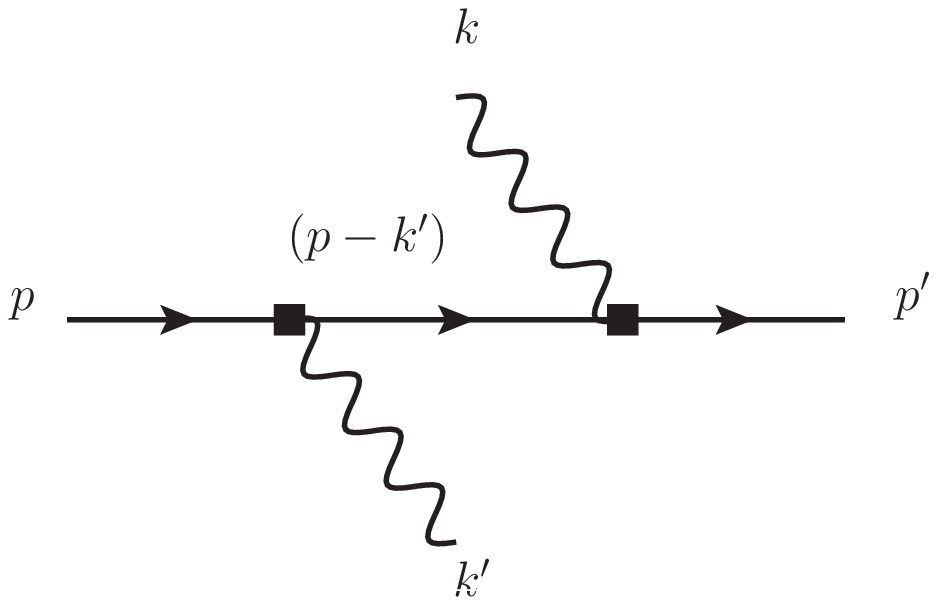}
}\vspace{-0.5cm}
\label{fig:coordinates}
\end{minipage}
&\quad
\begin{minipage}[c]{50mm}
\centerline{
\includegraphics[width=40mm, height=40mm, angle=0]{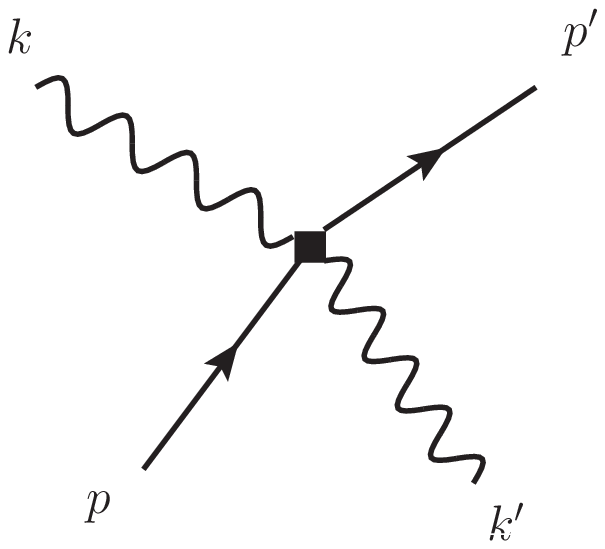}
}\vspace{-0.5cm}
\label{fig:localframe}
\end{minipage}
\end{tabular}
\vspace{0.5cm}
\caption[]{Feynman diagrams for Compton scattering in NCQED. }
\end{minipage}
\end{center}
\end{figure}
Here $u(p),{\overline u}({p^{'}}) $ denote
the spinor for incoming electron and outgoing electron respectively. $ \epsilon_{r_\nu}(k),\epsilon_{r^{'}\mu}^{*}(k^{'})$ are
polarization vectors for incoming photon(polarization r) and outgoing photon(polarization $r^{'}$) respectively. 
$H_{R,L}( \frac{1 \pm \gamma_5}{2})$ are 
projection operators for right and left helicity eigenstates of the electron. \\

Thus the total amplitude for the above process is given by the expression

\bea
{\mathcal{M}} &=& i {\mathcal{M}}_a +i {\mathcal{M}}_b +i {\mathcal{M}}_c \nonumber \\
 &=& ({\mathcal{M}}_{1a} + {\mathcal{M}}_{1b}+ {\mathcal{M}}_{1c})\epsilon_{r^{'}}^{*}(k^{'}) \epsilon_{r}(k) \nonumber \\
 &=& \{ -e^2(T_{11}^{\mu\nu} + T_{21}^{\nu\mu} ) + \frac{i e^2}{2}((T_{12}^{\Theta} + T_{13}^{\Theta})^{\mu\nu}+(T_{22}^{\Theta} + T_{23}^{\Theta})^{\nu\mu}- i(T^{\Theta})^{\mu\nu}) + O(\Theta^2) \}\epsilon_{r^{'}\mu}^{*}(k^{'}) \epsilon_{r_\nu}(k) \nonumber \\
\eea

The expressions of  various T's are given in Appendix A.

It is clear from above expression that interference 
between SM and NC terms can provide $O(\Theta)$ corrections to cross section. However one will
have to compensate for the imaginary factor i to get non vanishing $O(\Theta)$ correction
in cross section. This can be done by taking initial beam polarized which will then
generate a factor $i \epsilon^{\mu\nu\lambda\sigma}$ in Dirac traces and thus will produce non vanishing NC effects at leading order.

Since non commutative parameter is
considered as fundamental constant in nature, so its direction is fixed in some non rotating coordinate 
system(can be taken to be celestial sphere). However the experiment is done in laboratory coordinate system which 
is rotating with earth's rotation. So one should take into account these rotation effects on $\Theta^{\mu\nu}$ 
in this frame before moving towards the phenomenological investigations.

\begin{figure}[t]
\begin{center}\lvm
\begin{minipage}[c]{160mm}
\begin{tabular}{ccc}
\begin{minipage}[c]{80mm}
\centerline{
\includegraphics[width=80mm, height=60mm, angle=0]{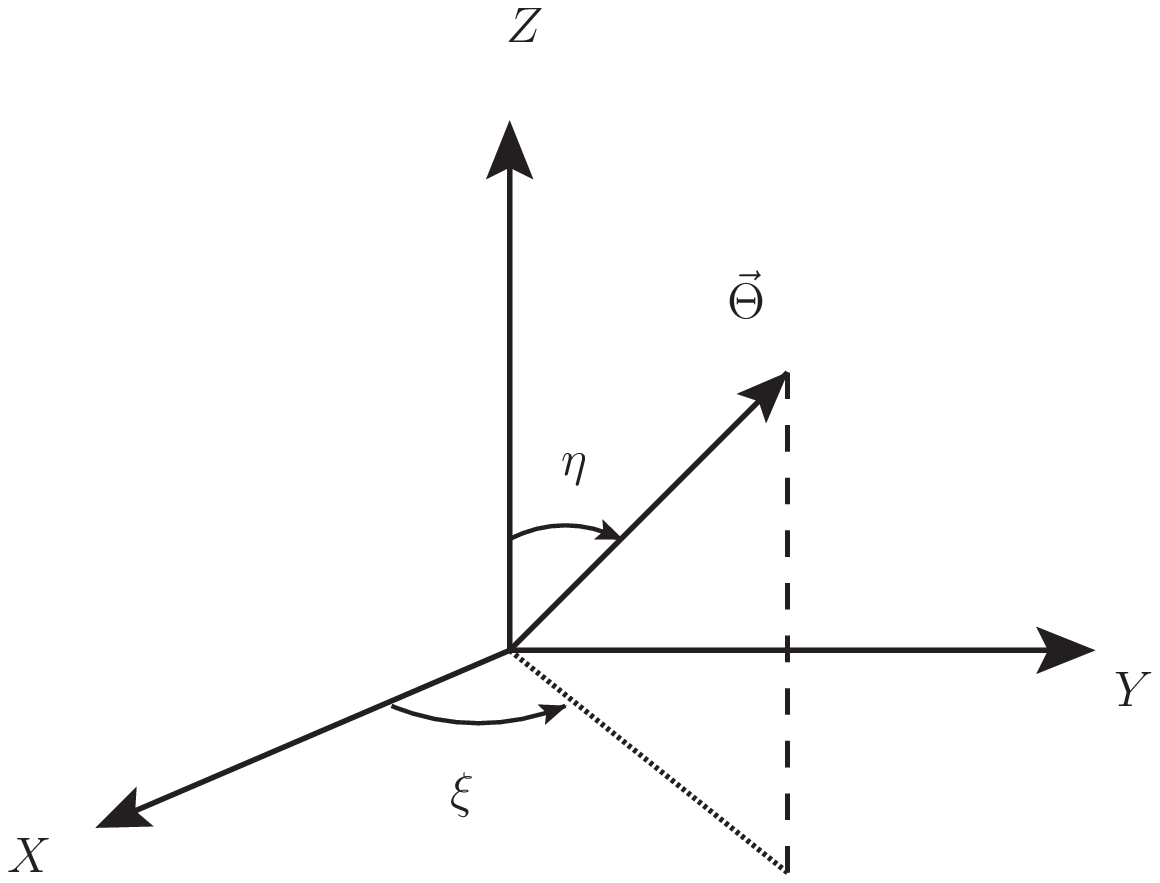}
}\vspace{-0.5cm}
\label{fig:coordinates}
\end{minipage}
&\quad
\begin{minipage}[c]{80mm}
\centerline{
\includegraphics[width=80mm, height=60mm, angle=0]{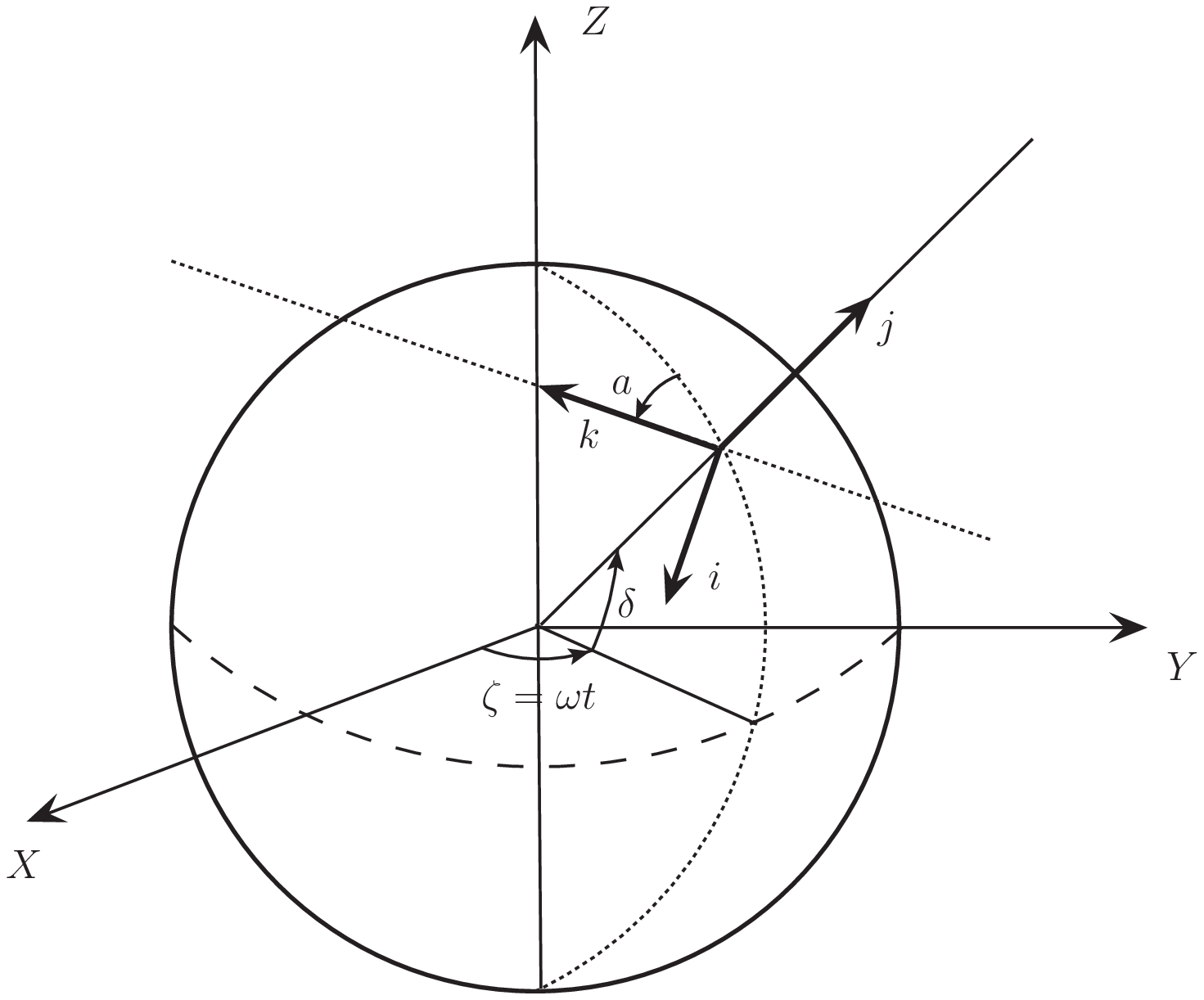}
}\vspace{-0.5cm}
\label{fig:localframe}
\end{minipage}
\end{tabular}
\vspace{0.5cm}
\caption[]{X-Y-Z is the primary coordinate system while $\{ \hat{i}-\hat{j}-\hat{k} \}$ are unit vectors pertaining  
to the laboratory coordinate system. The direction of $\vec{\Theta}$ is defined by angles $\eta$ and $\xi$.}
\end{minipage}
\end{center}
\end{figure}

These effects were considered in many previous studies\cite{earthroteffs1, earthroteffs2, earthroteffs3, earthroteffs4} but we shall follow reference\cite{earthroteffs3}.
In the laboratory coordinate system,
 the orthonormal basis of 
 the non rotating(primary) coordinate system($\hat{i}_X-\hat{j}_Y-\hat{k}_Z$) can be written as(see Fig.4)
\begin{eqnarray}
\begin{array}{ccc}
\hat{i}_X= 
   \left(\begin{array}{c}
      c_a s_\zeta + s_\delta s_a c_\zeta \\
      c_\delta c_\zeta \\
      s_a s_\zeta - s_\delta c_a c_\zeta 
   \end{array}\right),
&
\hat{j}_Y=
   \left(\begin{array}{c}
     -c_a c_\zeta + s_\delta s_a s_\zeta \\
      c_\delta s_\zeta \\
     -s_a c_\zeta - s_\delta c_a s_\zeta 
   \end{array}\right),
&
\hat{k}_Z=
   \left(\begin{array}{c}
      -c_\delta s_a  \\
       s_\delta  \\
       c_\delta c_a
   \end{array}\right).
\end{array}
\label{eqn:basis}
\end{eqnarray}\\
Here we have used the abberivations $c_\alpha = \cos \alpha, s_\alpha = \sin \alpha$ etc. ($\delta, a $) defines the location 
of experiment with $-\pi/2 \leq \delta \leq \pi/2$
and $0 \leq a \leq 2\pi$. Due to earth's rotation angle $\zeta$ increases with time and detector
comes to its original position after a cycle of one day so one can define 
$ \zeta = \omega t $ with $\omega = 2\pi/T_{day}$ where $ T_{day}=23h56m4.09053s $.
 
Thus the NC parameter in the Laboratory frame is given by electric and magnetic components
\bea
\vec{\Theta}_E &=& \Theta_E( \sin \eta_E \cos \xi_E ~\hat{i}_X +\sin\eta_E \sin\xi_E  ~\hat{j}_Y + \cos\eta_E ~\hat{k}_Z )\nonumber\\
\vec{\Theta}_B &=& \Theta_B(\sin \eta_B \cos \xi_B ~\hat{i}_X +  \sin\eta_B \sin\xi_B ~\hat{j}_Y+ \cos\eta_B ~\hat{k}_Z )\nonumber\\
\eea

with

\[ \vec{\Theta}_E = (\Theta^{01},\Theta^{02},\Theta^{03}) \hspace{1cm} \vec{\Theta}_B = (\Theta^{23},\Theta^{31},\Theta^{12})\]

and

\[ \Theta_E = |\vec{\Theta}_E|= 1/\Lambda_E^2 \hspace{1cm}  \Theta_B = |\vec{\Theta}_B|= 1/\Lambda_B^2 \]

Here  ($\eta, \xi$) specifies the direction of NC parameter($\Theta^{\mu\nu}$) w.r.t primary coordinates system with $0 \leq \eta \leq \pi$, $0 \leq \xi \leq 2\pi $.  
$\Theta_E$ and $\Theta_B$ are absolute values of its electric and magnetic components
with corresponding scales $\Lambda_E$ and $\Lambda_B$ respectively. One in general can probe them separately in
different processes.

Using these  definitions one can evaluate the various cross section observables with either standard Trace technique or by
helicity amplitude method. We follow here the Trace technique and various traces in cross sections are
evulated by using the Mathematica Package FeynCalc\cite{feyncalc}. The results are  cross checked in symbolic manipulation program FORM\cite{FORM}. Few details of the calculation are given in appendix A.\\

Thus in the Center of Mass frame ($A(p) + B(k) --> A(p^{'}) + B(k^{'})$)
\bea
p^{\mu} &=& \frac{\sqrt{s}}{2}\{1,0,0, 1\}\nonumber\\
k^{\mu} &=& \frac{\sqrt{s}}{2}\{1,0,0, -1\}\nonumber\\
{p^{\mu}}^{'} &=& \frac{\sqrt{s}}{2}\{1,\sin\theta\cos\phi, \sin\theta \sin\phi, \cos\theta\}\nonumber \\
{k^{\mu}}^{'} &=& \frac{\sqrt{s}}{2}\{1,-\sin\theta\cos\phi, -\sin\theta \sin\phi, -\cos\theta\}\nonumber \\
\eea

where $\theta$ is the polar angle and $\phi$ is the azimuthal angle, with initial beam direction
chosen as the z-axis.

The time dependence in cross section  enters through the NC parameter $\vec{\Theta}$
which changes with change of angle $\zeta$ because of earth's rotation. The final cross section formulae for different cases are given by:

{\bf{For Compton Scattering:}} \\
The differential cross section with keeping only incoming electron beam
in Right polarized state is given by

\begin{equation}
\displaystyle
\left(\frac{d\sigma}{d\Omega    }\right)_{\overrightarrow{\Theta_B}} =
    {\frac{\alpha^2}{8 s}}
\left[
      (2 \cos\theta +\cos^2\theta+5)\sec^2\frac{\theta}{2} + \bar{s}_B \{L_1^{\theta}( \Theta^{23}\cos\phi + \Theta^{31}\sin\phi )
+ L_2^{\theta} \Theta^{12}\}
\right]  \; ,
    \label{diff c.s.}
\end{equation}

where
\bea
\bar{s}_B &=& \frac{s}{\Lambda_B^2}, \hspace{1cm} L_1^{\theta}= 4\sin^2\frac{\theta}{2}(1 + \cos^2\frac{\theta}{2}) \tan\frac{\theta}{2};\hspace{1cm} L_2^{\theta}=4 \sin^2\frac{\theta}{2}
(1+\cos^2\frac{\theta}{2})\nonumber\\
\eea

Now the cross sections for Pair annihilation and Pair Creation are easy to evaluate since they are related
to the Compton by crossing symmetry. The cross section for Pair annihilation can be calculated by substituting $p^{'}\rightarrow -p^{'}, k \rightarrow -k$ while for Pair production can be found by inserting
$p \rightarrow -p, k^{'}\rightarrow -k^{'}$ in Compton trace expressions. The differential  cross section formulae for these two cases 
are given as:

{\bf{For Pair Annihilation:}} \\
The differential cross section with keeping only incoming electron beam
in Right polarized state is given by

\bea
\displaystyle
\left(\frac{d\sigma}{d\Omega    }\right)_{\overrightarrow{\Theta_E}} &=&
    {\frac{\alpha^2}{ s}}
\left[
      (1 +\cos^2\theta)\csc^2\theta  - \bar{s}_E \{ M_1^{\theta}(\Theta^{02} \cos\phi -  \Theta^{01}\sin\phi) \}
\right]  \; ,
    \label{diff c.s.}\nonumber\\
\eea

{\bf{For Pair Production:}}\\ 
Finally the differential cross section for  incoming $\gamma$ beams
in (L, R) state is given by

\bea
\displaystyle
\left(\frac{d\sigma}{d\Omega    }\right)_{\overrightarrow{\Theta_E}} &=&
    {\frac{2\alpha^2}{ s}}
\left[
      (1 +\cos^2\theta)\csc^2\theta - \bar{s}_E \{ N_1^{\theta}(\Theta^{02} \cos\phi -  \Theta^{01}\sin\phi) \}
\right]  \; ,
    \label{diff c.s.}\nonumber\\
\eea

\begin{equation}
\bar{s}_E = \frac{s}{\Lambda_E^2} \hspace{1cm} M_1^{\theta}=\cot\theta \hspace{1cm} N_1^{\theta}=
\frac{\csc\theta}{2}(1+\cos^2\theta) \nonumber\\
\end{equation}

The pure QED results can be recovered from the above expressions in the 
$\Lambda_E,\Lambda_B \rightarrow \infty $ limit.
Since it is difficult to get time dependent data,  we average over full day to be compared with
the experiment. We will use following cross section observables to examine the effects of non commutativity

\begin{eqnarray}
  \left\langle\frac{d{\sigma}}{d\cos\theta{d\phi}}\right\rangle_{T}
  &\equiv&
  \frac{1}{T_{day}}
  \int^{T_{day}}_{0}\frac{d\sigma^{}}{d\cos\theta{d\phi}}dt,
\\
  \left\langle\frac{d{\sigma}}{d\cos\theta}\right\rangle_{T}
  &\equiv&
  \frac{1}{T_{day}}
      \int^{T_{day}}_{0}\frac{d\sigma^{}}{d\cos\theta}dt,
\\
  \left\langle\frac{d{\sigma}}{d\phi}\right\rangle_{T}
  &\equiv&
  \frac{1}{T_{day}}\int^{T_{day}}_{0}\frac{d\sigma^{}}{d\phi}dt,
\label{eqn:dcr_dphi}
\\
  \langle\sigma\rangle_{T} &\equiv&
  \frac{1}{T_{day}}\int^{T_{day}}_{0}\sigma^{}dt,
\end{eqnarray}
where
\begin{eqnarray}
\frac{d{\sigma}^{}}{d\cos\theta}&\equiv&
  \int^{2\pi}_{0}\!\!d\phi
      \frac{d\sigma^{}}{d\cos\theta{d\phi}},\\
\frac{d{\sigma}^{}}{d\phi}&\equiv&
  \int^{1}_{-1}\!\!d(\cos\theta)
      \frac{d\sigma^{}}{d\cos\theta{d\phi}},\\
{\sigma}^{}&\equiv&
  \int^{1}_{-1}\!\!d(\cos\theta)
  \int^{2\pi}_{0}\!\!d\phi
      \frac{d\sigma^{}}{d\cos\theta{d\phi}}.
\end{eqnarray}

However in the case of Pair annihilation, final state photons are identical, so one counts all the possible
final states by integrating only over $ 0 < \theta \leq \pi/2 $. 

The initial phase dependence(i.e. $\xi$) disappears   in time
averaged observables and thus one can easily deduce $\{ \Theta, \eta\}$ from them.


\section{Numerical Results}
\noindent In this section we will provide the numerical results of our investigation. In order to 
determine \{$\vec{\Theta}_E, \vec{\Theta}_B$\} in laboratory system we studied variation of time averaged and time dependent cross 
section observables on $\{\xi, \eta, \Lambda\}$. We fixed the initial beam energy at $\sqrt{s}(=E_{com}) = 800$ GeV. 
The position of Lab
system is fixed by taking $\delta = \pi/4$ and $a=\pi/4$.

The NC corrections in  Pair annihilation and Pair creation, the $\phi$ dependence only appears in 
the form of $\cos\phi$ and $\sin\phi$. So non
commutative effects can  be obtained only in azimuthal angle distribution cross section($d\sigma/d\phi$) 
since they disappear in other observables once we integrate over the full azimuthal
angle($0-2\pi$). Hence for studying other cross section observables of these two processes 
we applied a cut of ($0-\pi$) on azimuthal angle $\phi$.

Figs. 5-8 represents the variation of different cross section observables for Compton, Figs. 9-12 for 
Pair annihilation and Figs. 13-16 for Pair production case.

\subsection{Time Averaged Angular distributions}
Non commutativity of space time defined by Eq.\ref{NCST} breaks Lorentz invariance including rotational invariance 
around the beam axis. This will lead to dependence of cross section on azimuthal angle which
is absent in Standard Model. Thus non commutativity of space time can provide  clear and distinct signature in azimuthal
angular variation of cross sections.

In this section we will discuss the time averaged  azimuthal
($\left\langle d\sigma/d\phi \right\rangle_{T}$) and total cross section 
($\left\langle \sigma \right\rangle_{T}$) for different values of $\{ \Lambda, \eta\}$.
Our results are  useful for case $s/\Lambda^2 < 1$ since in this domain one can safely ignore
higher order corrections to cross section.

The angle $\eta$ can be determined  by fitting the shape of curve of $(d\sigma/d\phi)_T$ plotted for different
values of $\eta$ for a fixed value of non commutative scale especially around
$\phi = 2\pi/3, 5\pi/3$ in Compton and $\phi = 6\pi/5, \pi/5$ in Pair annihilation and Pair
production case where respectively there is maximum enhancement and deficit in cross section 
compared to pure
QED case. Similarly magnitude of NC parameter can be determined from fitting the curve of $(d\sigma/d\phi)_T$ plotted for different values of $\Lambda$ for a fixed $\eta$. For a 
fixed center of mass energy($\sqrt{s}$) the deviations to QED cross section becomes larger and larger as one lowers  the value of non commutative scale($\Lambda$).

Since NC corrections in time averaged cross sections for all three cases are proportional
to $\cos\eta$ so their effect becomes maximum at $\eta=0,\pi$. Also correction is equal and opposite in
magnitude for $\pi-\eta$ case. Thus polarization of beams is  more useful compared to unpolarized  case\cite{earthroteffs3}  where
there is a two fold ambiguity in determination of $\eta$ for Pair annihilation.

The time averaged total cross section is also sensitive to the $|\Theta|$ and $\eta$. Thus one can 
also infer the information about  the magnitude of NC parameter by fitting the ($\left\langle \sigma \right\rangle_{T}$)
curve plotted vs $\eta$ for different values of NC scale especially around $\eta =0, \pi$ where
respectively there is maximum enhancement and deficit for Compton while opposite for Pair annihilation and
Pair production.

Hence one can determine $\{\eta_B, \Theta_B \}$ from figures pertaining to Compton  while 
$\{\eta_E, \Theta_E \}$ can be
obtained from the Pair annihilation and Pair production curves without any ambiguity. 

\subsection{Time Dependent total Cross section}
In order to obtain angle $\xi$, we have studied the time variation of total cross section vs $\omega t-\xi$
with different values of $\eta$. 

Figs. 8,12,16 gives the variation of $\sigma$ vs $\omega t-\xi$ for $\Lambda = 1$ TeV with different values of $\eta$.
It is clear from them that the cases with different $\eta$ are clearly distinguishable from each other.
If the time variation of total cross section is observed then we can determine the magnitude and direction of $\{\vec{\Theta} \}$  from the curves in terms of three parameters
$\{ \Theta, \eta, \xi \}$. The $\{\Theta, \eta \}$ can be obtained by fitting the magnitude and 
shape of curves  while $\xi$ can be determined from the phase of time evolution of $\sigma$.

Although $\xi$ can be determined from the time variation of differential cross sections instead of
total cross section however one can imagine that  to plan such type of experiment needs very large luminosity because we must divide not only the
phase space but also the time distribution into many bins, in order to get such dependence.

This completes our discussion for the determination of direction and magnitude of electric and magnetic 
components ($\{\vec{\Theta}_E, \vec{\Theta}_B \}$) of NC parameter $\Theta$.



\section{Summary and Discussion}
The NCSM is one of the extension for Physics beyond SM with motivations 
from string theory and quantum gravity. Its phenomenological implications are
quite  interesting since scale of non commutativity could be as low 
as a few TeV, which can be explored at present or future colliders. 

In the present work we have investigated the TeV scale signatures of NC  space-time in $e^{+}e^{-}\rightarrow \gamma\gamma$, 
$e^{-}\gamma \rightarrow e^{-}\gamma$ and $\gamma\gamma \rightarrow e^{+}e^{-} $ processes. We have done our study with initial 
beam polarization effects which offers
the unique opportunity of having deviations from the SM cross sections occur at O$(\Theta)$. Previous
studies are mainly done with unpolarized case where these effects appear at O$(\Theta^2)$. In this 
analysis we have also taken into account the apparent time variation
of non commutative parameter($\Theta^{\mu\nu}$) in Laboratory frame. The primary coordinate system
is fixed to the celestial sphere.

The NC corrections to Compton are sensitive to 
the magnetic component($\vec{\Theta}_B$) while for Pair production and Pair annihilation can probe 
electric component($\vec{\Theta}_E$) of NC parameter ($\vec{\Theta}$). Since such theories breaks  rotational invariance 
around the beam axis, it leads to dependence of cross section on azimuthal angle which
is absent in Standard Model. Thus they can provide clear and completely distinguishable
signatures in azimuthal angular variation of cross sections. 

To determine $\vec{\Theta}$ we have studied variation of various cross sections observables.
Magnitude  $|\vec{\Theta}|$ and angle $\eta$ can be determined by fitting the shape of curves of ($d\sigma/d\phi$) plotted for different values of $\eta$ and $\Lambda$ respectively. From time variation of total cross section one can determine magnitude as well as direction of NC parameter i.e. $\xi$, $\eta$ and $|\vec{\Theta}|$. These implications of non commutative space time can  be tested at proposed  International Linear Collider(ILC). 

In this study for illustration purposes Lab coordinates are taken to be $(\delta, a)=(\pi/4, \pi/4)$. However in experiments with several
detector sites such as LEP, the direction of incoming beam in each site is set
to be along the different direction. Then angular distributions as well as
time variation of observables will behave differently at each point because of the difference in direction
of $\vec{\Theta}$ at different interacting points. Therefore
combined analysis of various results from several experiments at different locations can help in 
probing the non commutative nature of space time.

\begin{acknowledgments}
S.K.G is grateful to Dr. Rolf Mertig for useful correspondence and Prof. B. Ananthanarayan
for discussions. The work of N.G.D is supported by
the US DOE under Grant No. DE-FG02-96ER40969.
 The work of P.K.D is supported by the DST Fast Track project SR/FTP/PS-11/2006 and
BITS Seed Grant project, 2011. 
 S.K.G acknowleges partial support from DST Ramanujan Fellowship
SR/S2/RJN-25/2008. T.Shreecharan thanks UGC India for financial support through 
their Dr. D. S. Kothari
Post-Doctoral Fellowship Scheme.\\

\end{acknowledgments}

\section{Appendix A: Compton Scattering}

In this appendix we will reveal some details of our calculation. The complete $O(\Theta)$
Feynman amplitude square for Compton scattering is given by the expression:

\bea
{\mathcal{M}} {\mathcal{M}}^{\dag} &=&  ({\mathcal{M}} {\mathcal{M}}^{\dag})^{CM} + ({\mathcal{M}} {\mathcal{M}}^{\dag})^{NC}_{CM}  \nonumber \\
 &=& e^4 \{ T_{11} T_{11}^{\dag} + T_{21} T_{21}^{\dag} + T_{21} T_{11}^{\dag} + T_{11} T_{21}^{\dag}  \}\epsilon_{r^{'}}^{*}(k^{'}) \epsilon_{r }(k) 
\epsilon_{r^{'} }(k^{'}) \epsilon_{r }^{*}(k)\nonumber \\
&+& \frac{i e^4}{2}\{ T_{11}({T_{12}^{\Theta}}^{\dag} + {T_{13}^{\Theta}}^{\dag})  + T_{21}({T_{22}^{\Theta}}^{\dag} + {T_{23}^{\Theta}}^{\dag}) + T_{11}({T_{22}^{\Theta}}^{\dag} + {T_{23}^{\Theta}}^{\dag})
+ T_{21}({T_{12}^{\Theta}}^{\dag} + {T_{11}^{\Theta}}^{\dag})\nonumber\\
&+& i T^{\Theta}({T_{11}}^{\dag} + {T_{21}}^{\dag}) -(T_{12}^{\Theta} + T_{13}^{\Theta}) T_{11}^{\dag} - (T_{22}^{\Theta} + T_{23}^{\Theta}) T_{21}^{\dag} - (T_{22}^{\Theta} + T_{23}^{\Theta}) T_{11}^{\dag} \nonumber\\
&-& (T_{12}^{\Theta} + T_{13}^{\Theta}) T_{21}^{\dag} +i (T_{11} + T_{21}) {T^{\Theta}}^{\dag} \}\epsilon_{r^{'}}^{*}(k^{'}) \epsilon_{r }(k) 
\epsilon_{r^{'} }(k^{'}) \epsilon_{r }^{*}(k)\nonumber\\
&+& O(\Theta^2) terms +..........
\eea

Here CM denotes the pure commutative part and NC-CM is $O(\Theta)$corrected terms arising due to interference between
Commutative and Non commutative part. 
The various 
involved T terms are given by

\bea
({T}_{11})^{\mu\nu} &=& \left[{\overline u}({p^{'}}) H_{(L,R)} \gamma^\mu \frac{i}{\psla+\ksla} \gamma^\nu u(p) \right] \nonumber \\
({T}_{12}^{\Theta})^{\mu\nu} &=& \left[{\overline u}({p^{'}}) H_{(L,R)} (\Gamma_{12}^{\Theta})^\mu \frac{i}{\psla+\ksla} \gamma^\nu u(p) \right] \nonumber \\
({T}_{13}^{\Theta})^{\mu\nu} &=& \left[{\overline u}({p^{'}}) H_{(L,R)} \gamma^\mu \frac{i}{\psla+\ksla} (\Gamma_{13}^{\Theta})^\nu u(p) \right] \nonumber \\
\eea

\bea
({T}_{21})^{\mu\nu} &=& \left[{\overline u}({p^{'}}) H_{(L,R)} \gamma^\mu \frac{i}{\psla-\kprsla} \gamma^\nu u(p) \right] \nonumber\\
({T}_{22}^{\Theta})^{\mu\nu} &=& \left[{\overline u}({p^{'}}) H_{(L,R)} (\Gamma_{22}^{\Theta})^\mu \frac{i}{\psla-\kprsla} \gamma^\nu u(p) \right] \nonumber \\
({T}_{23}^{\Theta})^{\mu\nu} &=& \left[{\overline u}({p^{'}}) H_{(L,R)} \gamma^\mu \frac{i}{\psla-\kprsla} (\Gamma_{23}^{\Theta})^\nu u(p) \right] \nonumber \\
\eea

\bea
({T}^{\Theta})^{\mu\nu} &=& \left[{\overline u}({p^{'}})\Theta^{\mu\nu\rho}(k^{'}+k)_{\rho} H_{(L,R)}  u(p) \right] \nonumber\\
\eea

where different $\Gamma^{\Theta}$ terms are given by

\bea
({\Gamma}_{13}^{\Theta})^{\mu} &=&  ((p+k)\Theta p)\gamma^{\mu}-((p+k)\Theta)^{\mu}\psla-(\psla + \ksla)(\Theta p)^{\mu} \nonumber\\
({\Gamma}_{12}^{\Theta})^{\mu} &=&  (p^{'}\Theta (p+k))\gamma^{\mu}-(p^{'}\Theta)^{\mu}(\psla + \ksla)-\psla^{'}(\Theta (p+k))^{\mu}  \nonumber \\
({\Gamma}_{23}^{\Theta})^{\mu} &=& ((p-k^{'})\Theta p)\gamma^{\mu}-((p-k^{'})\Theta)^{\mu}\psla-(\psla - \ksla^{'})(\Theta p)^{\mu} \nonumber \\
({\Gamma}_{22}^{\Theta})^{\mu} &=&  (p^{'}\Theta (p-k^{'}))\gamma^{\mu}-(p^{'}\Theta)^{\mu}(\psla - \ksla^{'})-\psla^{'} (\Theta (p-k^{'}))^{\mu}  \nonumber \\
\eea

Using these expressions one can put the different terms of ${\mathcal{M}} {\mathcal{M}}^{\dag} $ in form of trace expressions.
e.g.
\bea
T_{11} T_{11}^{\dag} &=& \frac{1}{(p + k)^{4}} Tr\{ \psla^{'} H_{(L,R)} \gamma^{\mu} (\psla + \ksla) \gamma^{\nu} \psla \gamma^{b} (\psla + \ksla) \gamma^a\} \nonumber \\
T_{21} T_{21}^{\dag} &=& \frac{1}{(p - k^{'})^{4}} Tr\{ \psla^{'} H_{(L,R)} \gamma^{\nu} (\psla - \ksla^{'}) \gamma^{\mu} \psla \gamma^{a} (\psla - \ksla^{'}) \gamma^b\} \nonumber \\
T_{12}^{\Theta} T_{11}^{\dag} &=& \frac{1}{(p + k)^{4}} Tr\{ \psla^{'} H_{(L,R)} (X_{12}^{\Theta})^{\mu} (\psla + \ksla) \gamma^{\nu} \psla \gamma^{b} (\psla + \ksla) \gamma^a\} \nonumber \\
T_{13}^{\Theta} T_{21}^{\dag} &=& \frac{1}{(p + k)^{2}(p - k^{'})^{2}} Tr\{ \psla^{'} H_{(L,R)} \gamma^{\mu} (\psla + \ksla) (X_{13}^{\Theta})^{\nu}  \psla \gamma^{a} (\psla - \ksla^{'}) \gamma^b\} \nonumber \\
\eea

with
\bea
(X_{12}^{\Theta})^{\mu} &=& C_2 \gamma^{\mu} - (V_3)^{\mu}(\psla + \ksla) +\psla^{'} (V_1)^{\mu} \nonumber\\
(X_{13}^{\Theta})^{\mu} &=& C_1 \gamma^{\mu} - (V_1)^{\mu}\psla -(\psla + \ksla)(V_2)^{\mu} \nonumber\\
\eea

Here 
\bea 
C_1 &=& (p + k)\Theta p;\hspace{0.5cm} C_2 = p^{'}\Theta(p+k) \nonumber\\
V_1 &=& (p+k)\Theta; \hspace{0.5cm}V_2 = \Theta p ; \hspace{0.5cm}V_3 = p^{'}\Theta \nonumber\\
\eea

Then these trace expressions can be evaluated using the 
Mathematica packages like FeynCalc\cite{feyncalc} or Symbolic Manipulation programme FORM\cite{FORM}. In this way one can obtain various cross section
observables.




\newpage
\begin{figure}[t]
\begin{center}\lvm
\begin{minipage}[c]{160mm}
\begin{tabular}{cc}
\begin{minipage}[c]{80mm}
\centerline{
\includegraphics[width=85mm, height=95mm, angle=0]{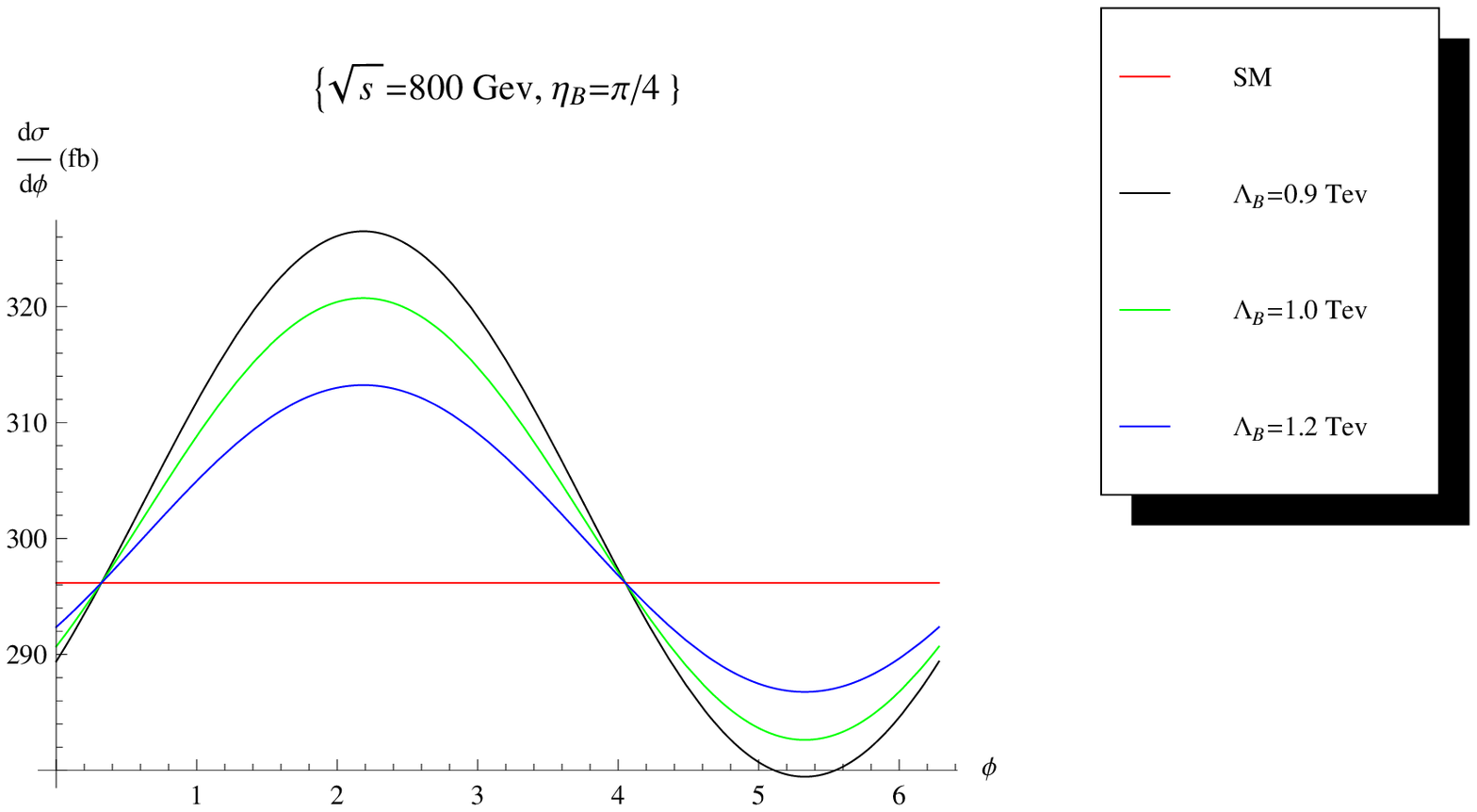}
}\vspace{-2cm}
\caption[]{Compton Scattering:~ Time averaged polar angle distribution vs 
$\phi$ for different values of NC scale($\Lambda_B$).
             }
\label{fig:coordinates}
\end{minipage}
&\quad
\begin{minipage}[c]{80mm}
\centerline{
\includegraphics[width=85mm, height=95mm, angle=0]{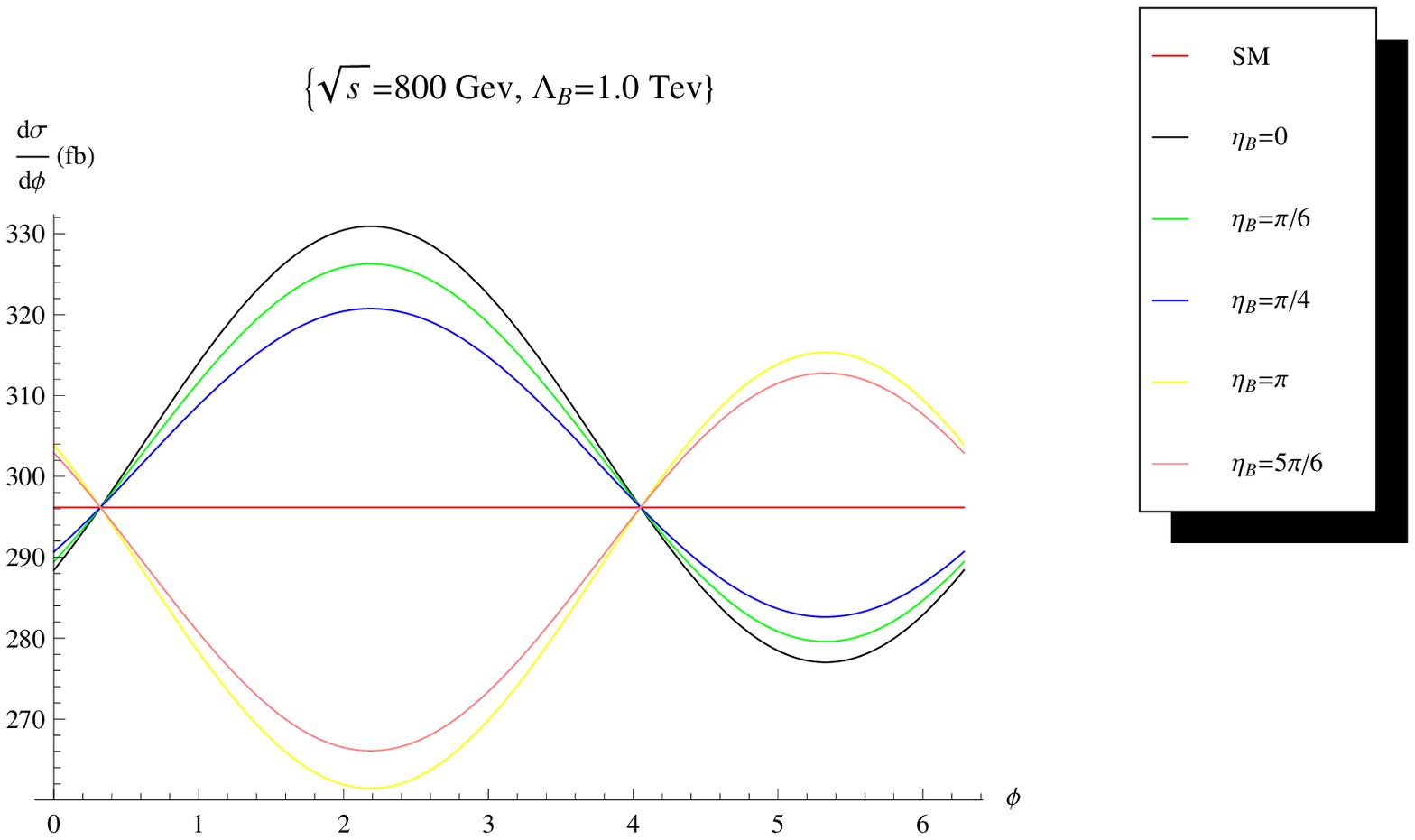}
}\vspace{-2cm}
\caption[]{Compton Scattering:~ Time averaged azimuthal angle distribution vs $\phi$
for different values of angle $\eta_B$.}
\label{fig:localframe}
\end{minipage}
\end{tabular}
\end{minipage}
\end{center}
\end{figure}

\begin{figure}[t]
\begin{center}\lvm
\begin{minipage}[c]{160mm}
\begin{tabular}{cc}
\begin{minipage}[c]{80mm}
\centerline{
\includegraphics[width=85mm, height=95mm, angle=0]{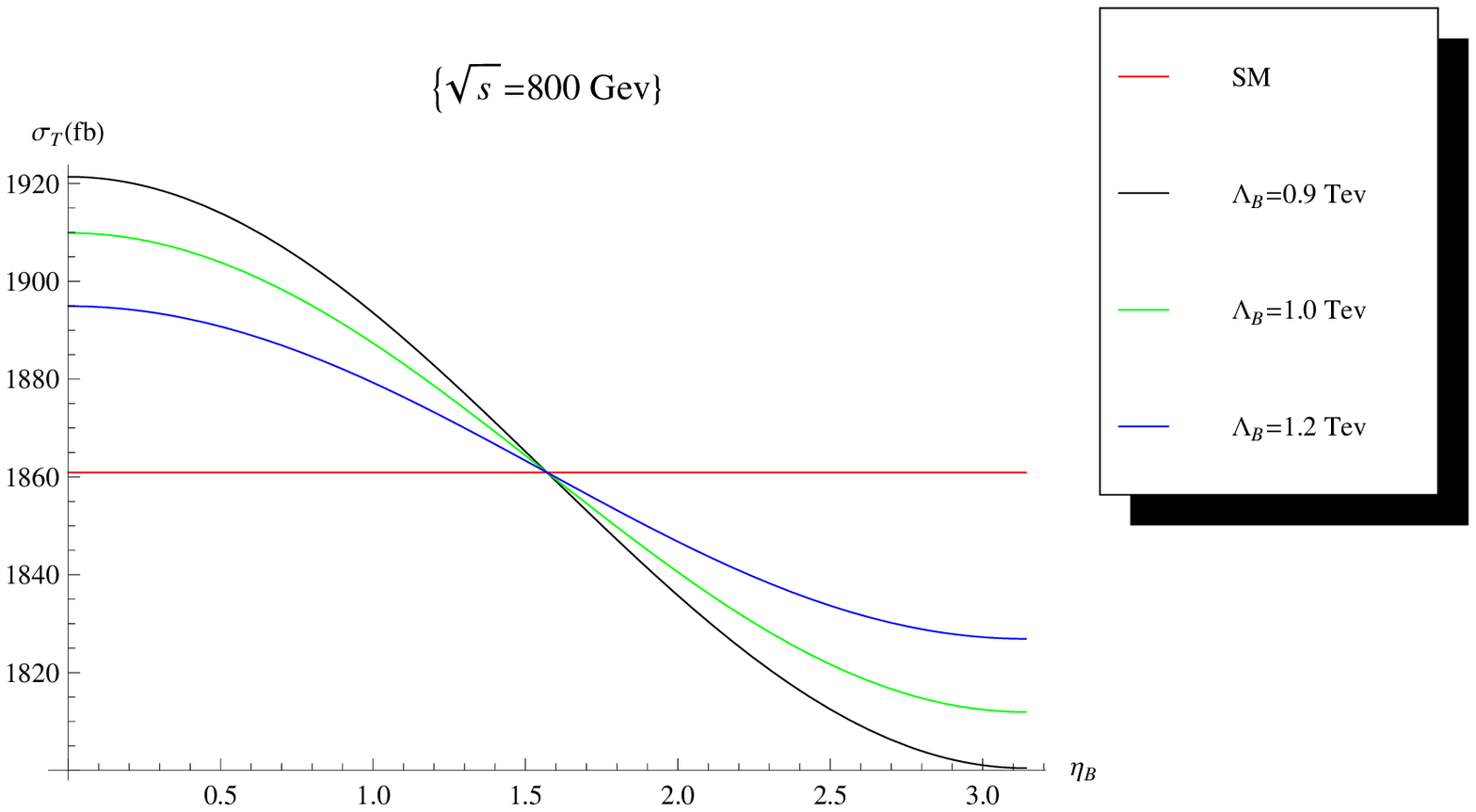}
}\vspace{-2cm}
\caption[]{Compton Scattering:~ Time average of total cross section $(\sigma_T)$ for different
values of NC scale($\Lambda_B$).
             }
\label{fig:coordinates}
\end{minipage}
&\quad
\begin{minipage}[c]{80mm}
\centerline{
\includegraphics[width=85mm, height=95mm, angle=0]{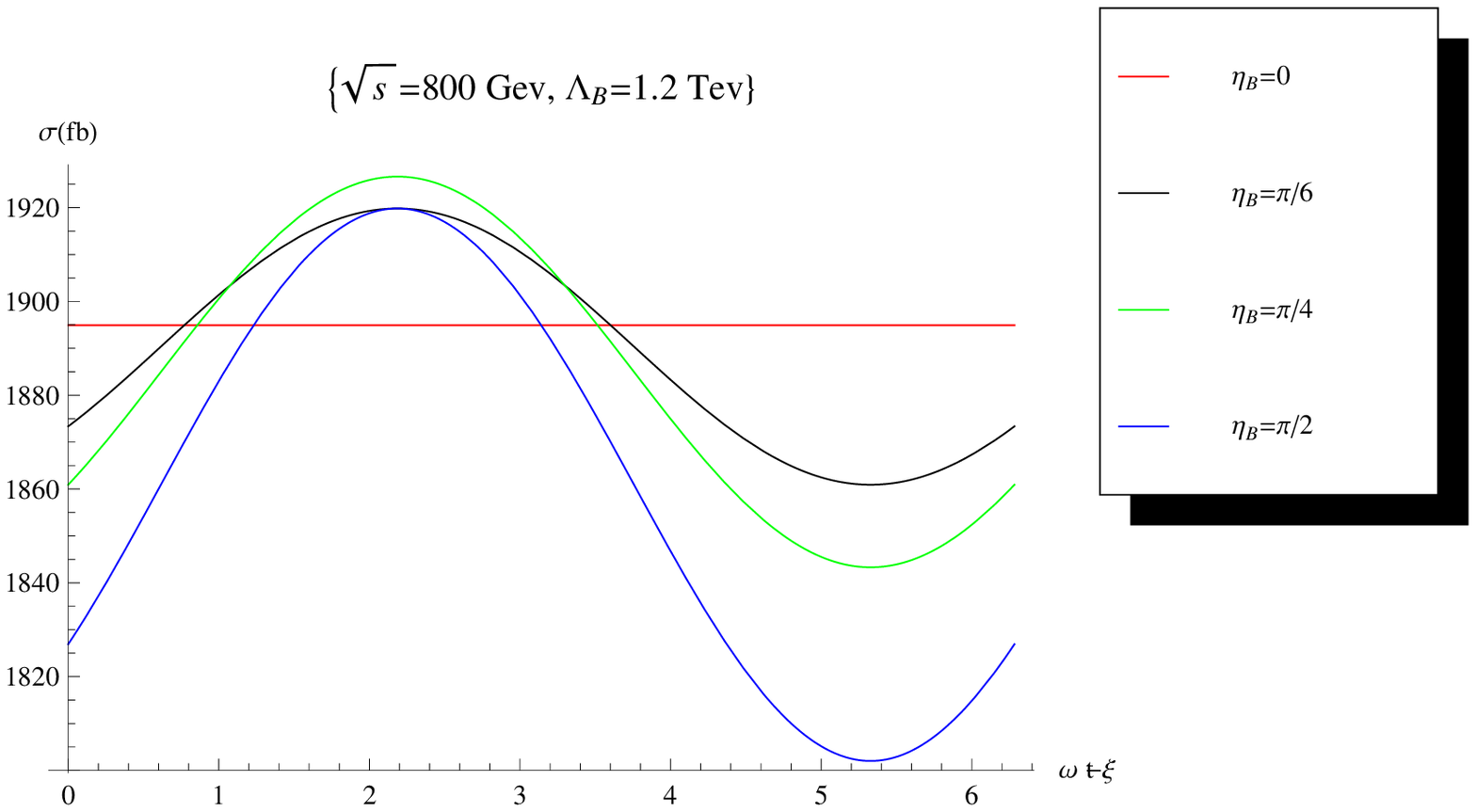}
}\vspace{-2cm}
\caption[]{Compton Scattering:~Time dependent total cross section $(\sigma)$ vs phase($\omega t -\xi$)
for different values of angle $\eta_B$.}
\label{fig:localframe}
\end{minipage}
\end{tabular}
\end{minipage}
\end{center}
\end{figure}

\newpage
\begin{figure}[t]
\begin{center}\lvm
\begin{minipage}[c]{160mm}
\begin{tabular}{cc}
\begin{minipage}[c]{80mm}
\centerline{
\includegraphics[width=85mm, height=95mm, angle=0]{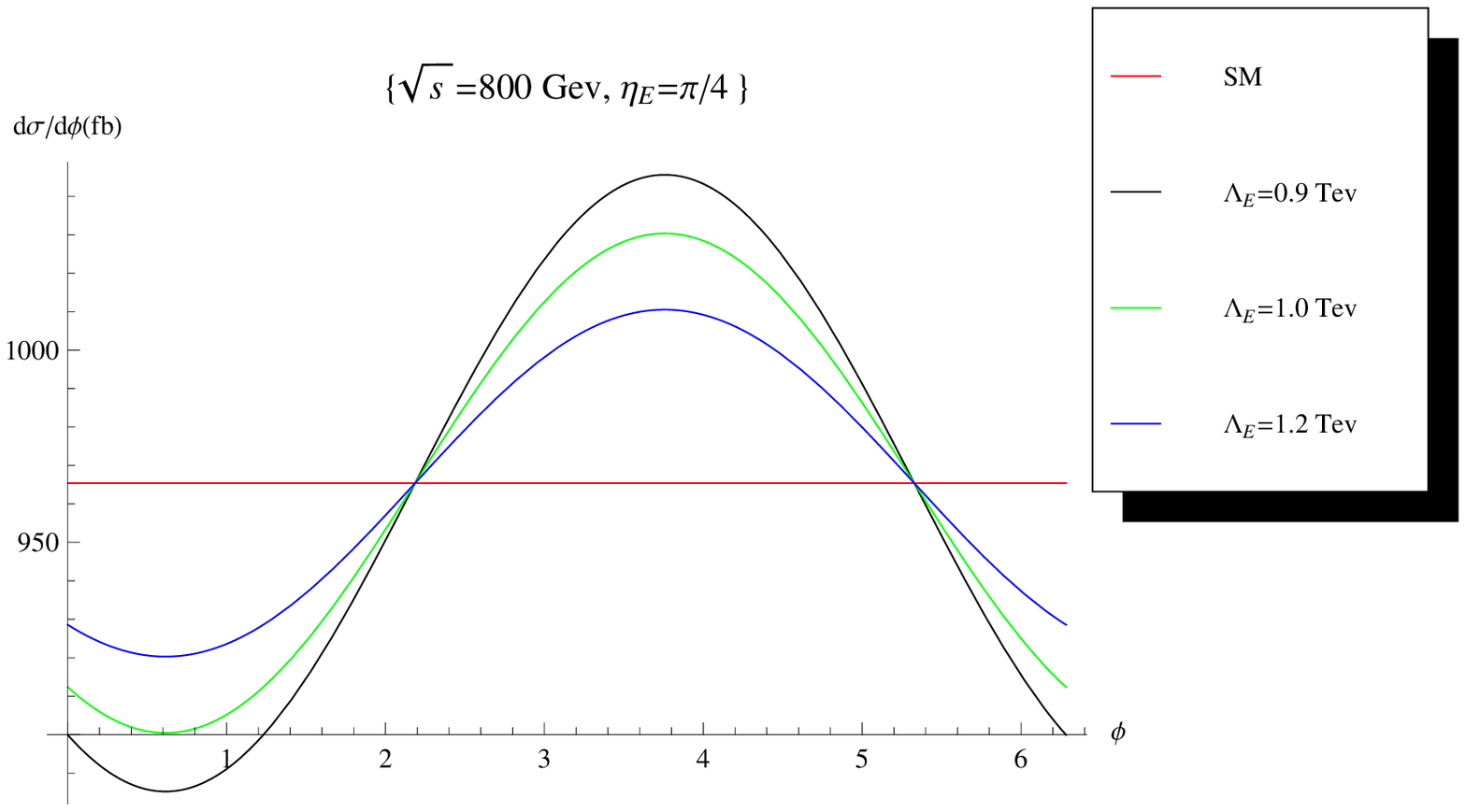}
}\vspace{-2cm}
\caption[]{Pair Production:~ Time averaged polar angle distribution vs 
$\phi$ for different values of NC scale($\Lambda_E$).
             }
\label{fig:coordinates}
\end{minipage}
&\quad
\begin{minipage}[c]{80mm}
\centerline{
\includegraphics[width=85mm, height=95mm, angle=0]{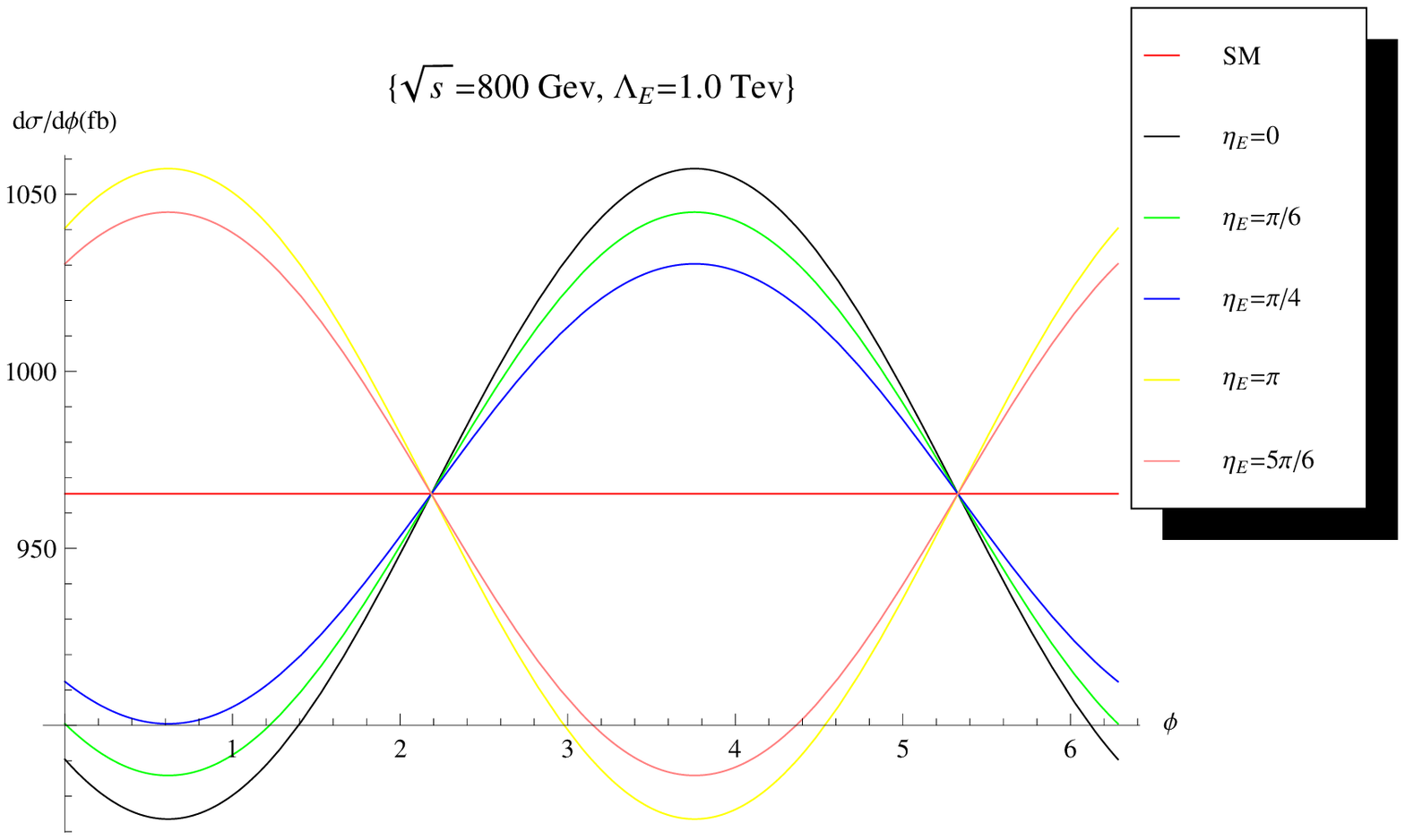}
}\vspace{-2cm}
\caption[]{Pair Production:~ Time averaged azimuthal angle distribution vs $\phi$
for different values of angle $\eta_E$.}
\label{fig:localframe}
\end{minipage}
\end{tabular}
\end{minipage}
\end{center}
\end{figure}

\begin{figure}[t]
\begin{center}\lvm
\begin{minipage}[c]{160mm}
\begin{tabular}{cc}
\begin{minipage}[c]{80mm}
\centerline{
\includegraphics[width=85mm, height=95mm, angle=0]{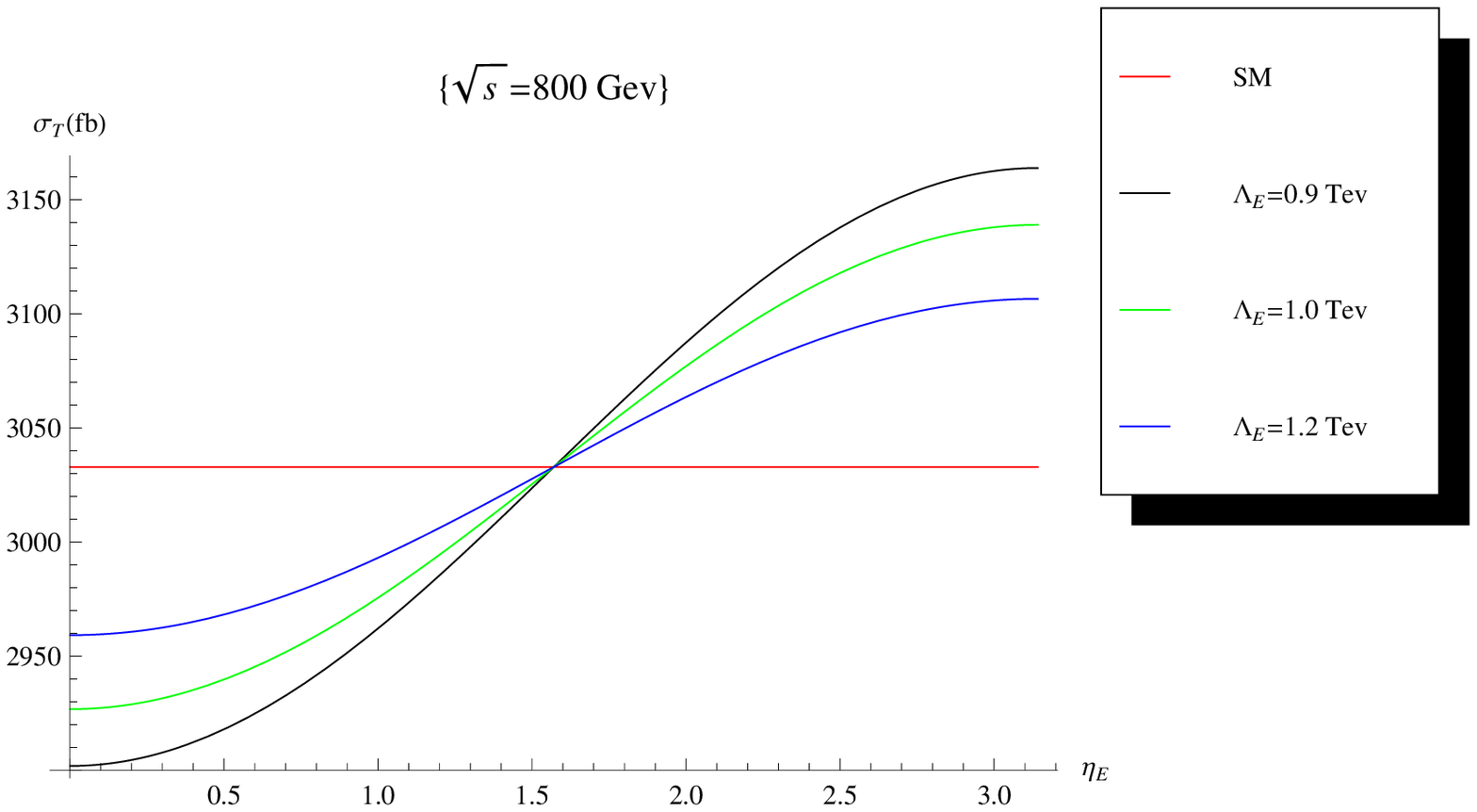}
}\vspace{-2cm}
\caption[]{Pair Production:~ Time average of total cross section $(\sigma_T)$ for different
values of NC scale($\Lambda_E$).
             }
\label{fig:coordinates}
\end{minipage}
&\quad
\begin{minipage}[c]{80mm}
\centerline{
\includegraphics[width=85mm, height=95mm, angle=0]{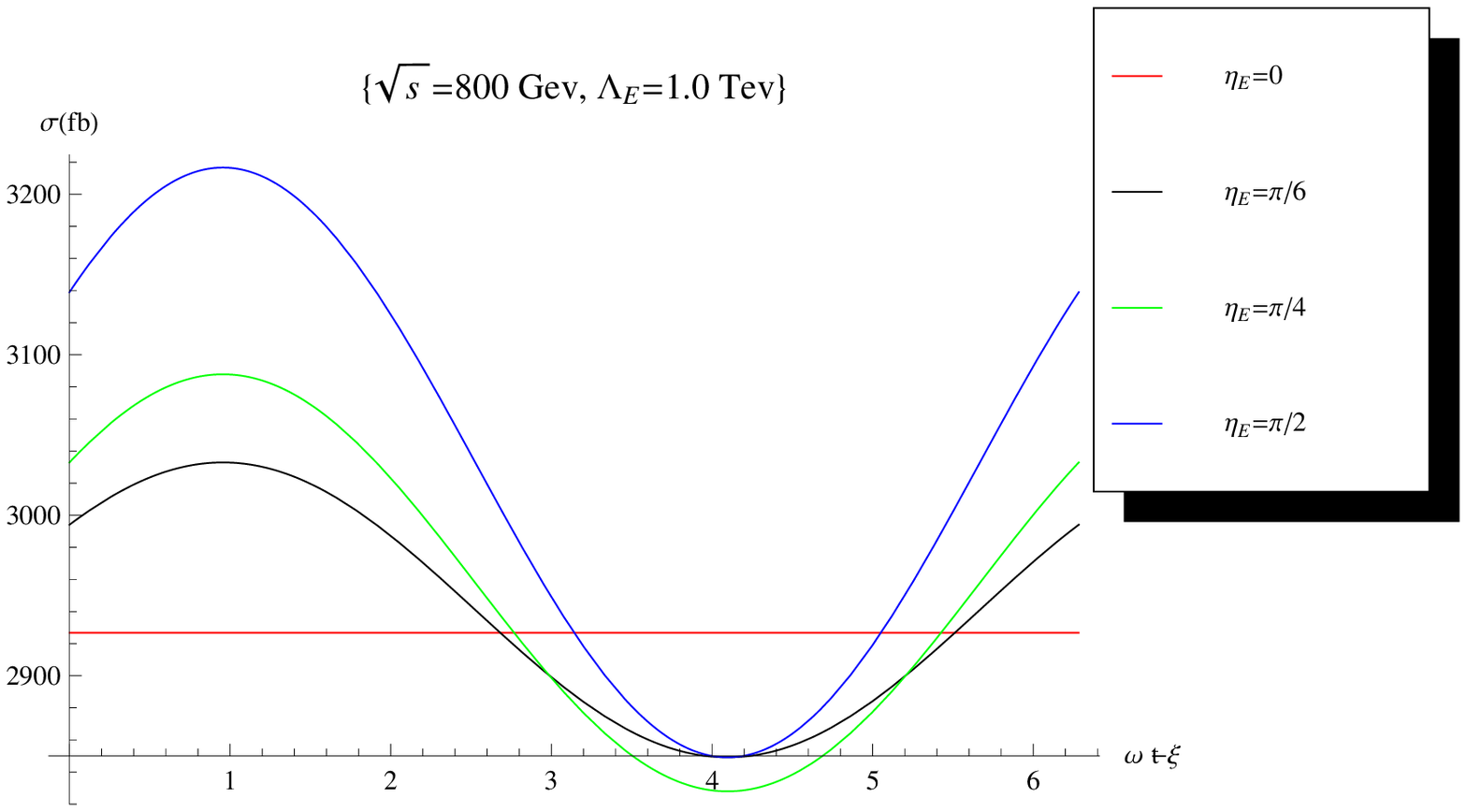}
}\vspace{-2cm}
\caption[]{Pair Production:~Time dependent total cross section $(\sigma)$ vs phase($\omega t -\xi$)
for different values of angle $\eta_E$.}
\label{fig:localframe}
\end{minipage}
\end{tabular}
\end{minipage}
\end{center}
\end{figure}

\newpage
\begin{figure}[t]
\begin{center}\lvm
\begin{minipage}[c]{160mm}
\begin{tabular}{cc}
\begin{minipage}[c]{80mm}
\centerline{
\includegraphics[width=85mm, height=95mm, angle=0]{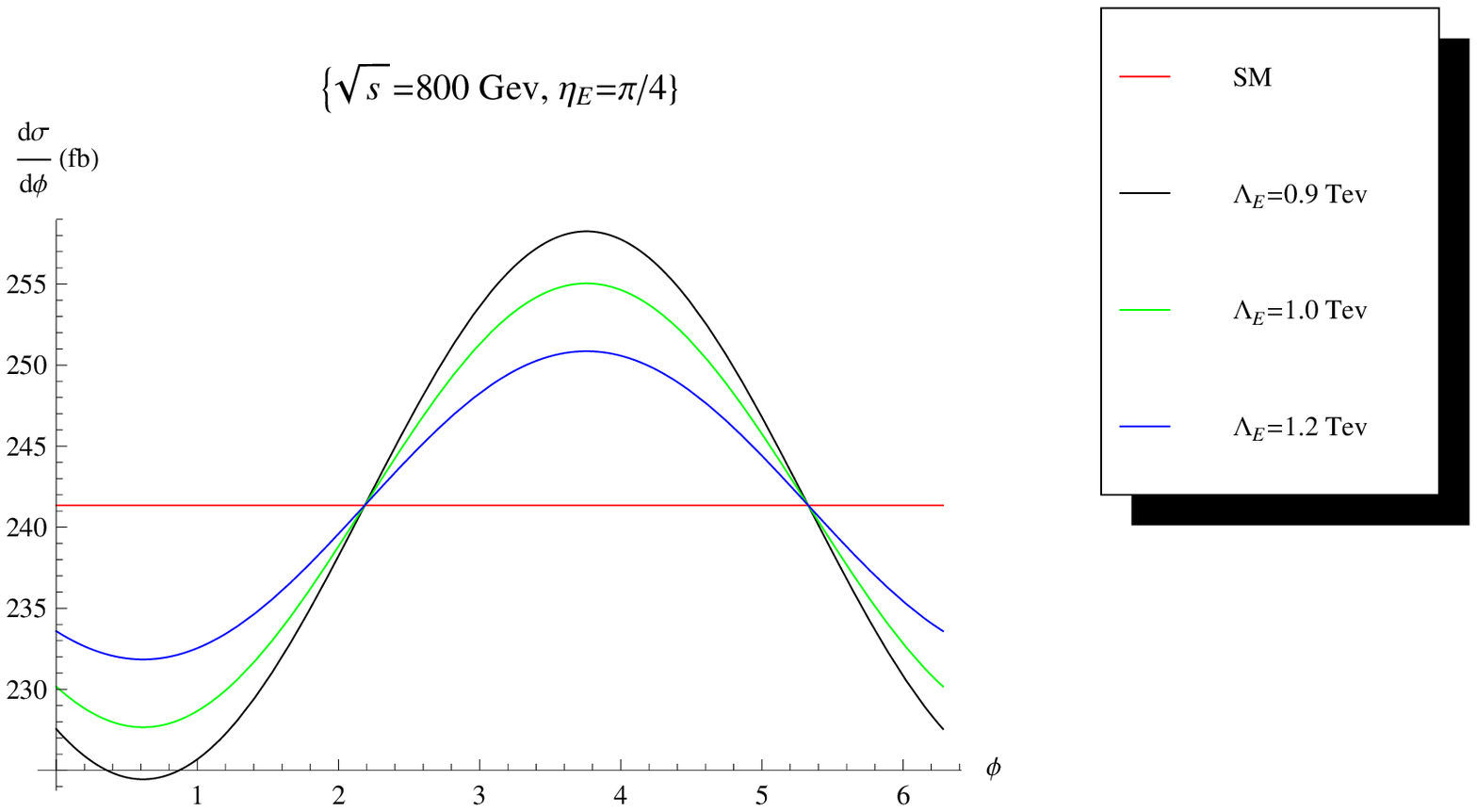}
}\vspace{-2cm}
\caption[]{Pair Annihilation:~ Time averaged polar angle distribution vs 
$\phi$ for different values of NC scale($\Lambda_E$).
             }
\label{fig:coordinates}
\end{minipage}
&\quad
\begin{minipage}[c]{80mm}
\centerline{
\includegraphics[width=85mm, height=95mm, angle=0]{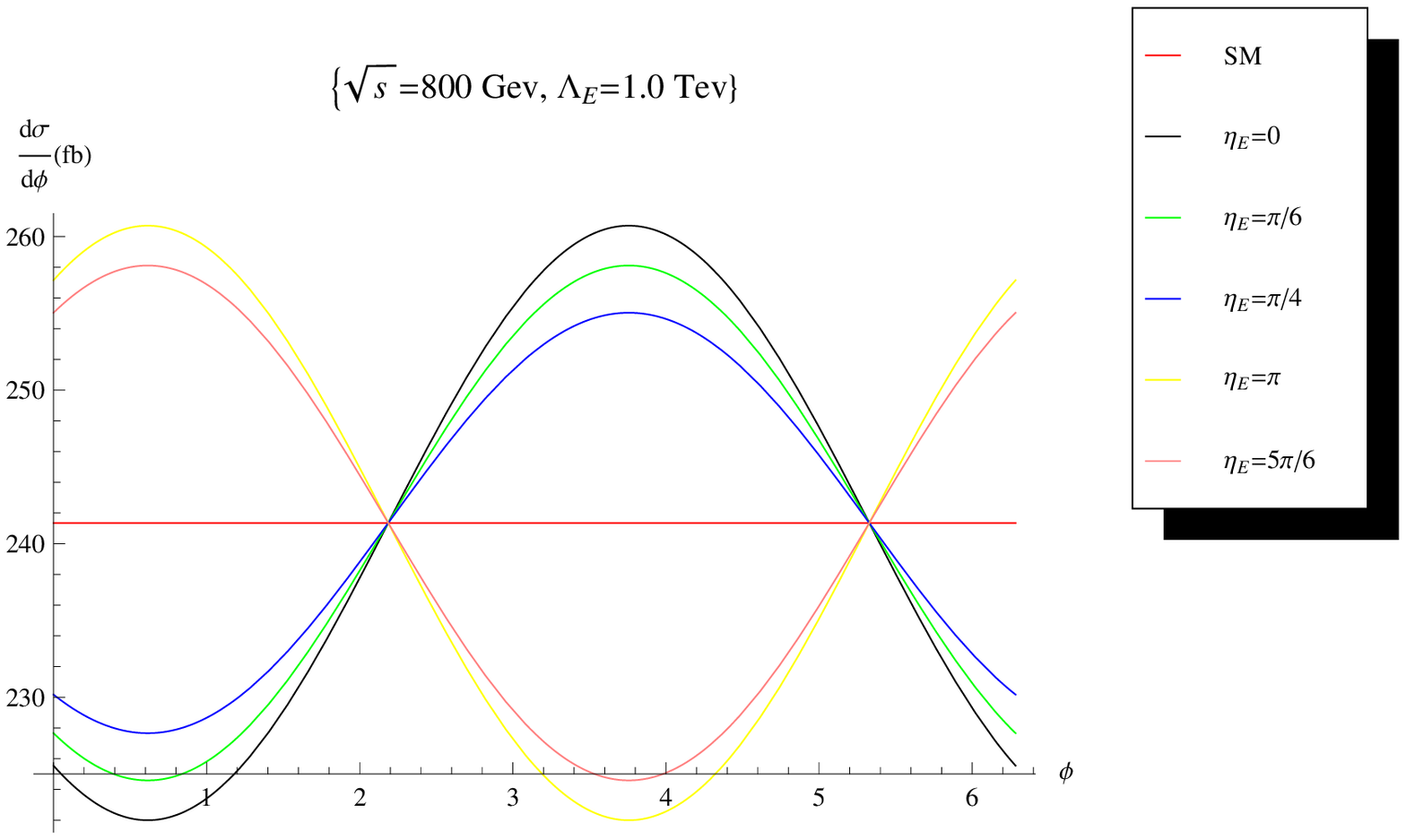}
}\vspace{-2cm}
\caption[]{Pair Annihilation:~ Time averaged azimuthal angle distribution vs $\phi$
for different values of angle $\eta_E$.}
\label{fig:localframe}
\end{minipage}
\end{tabular}
\end{minipage}
\end{center}
\end{figure}

\begin{figure}[t]
\begin{center}\lvm
\begin{minipage}[c]{160mm}
\begin{tabular}{cc}
\begin{minipage}[c]{80mm}
\centerline{
\includegraphics[width=85mm, height=95mm, angle=0]{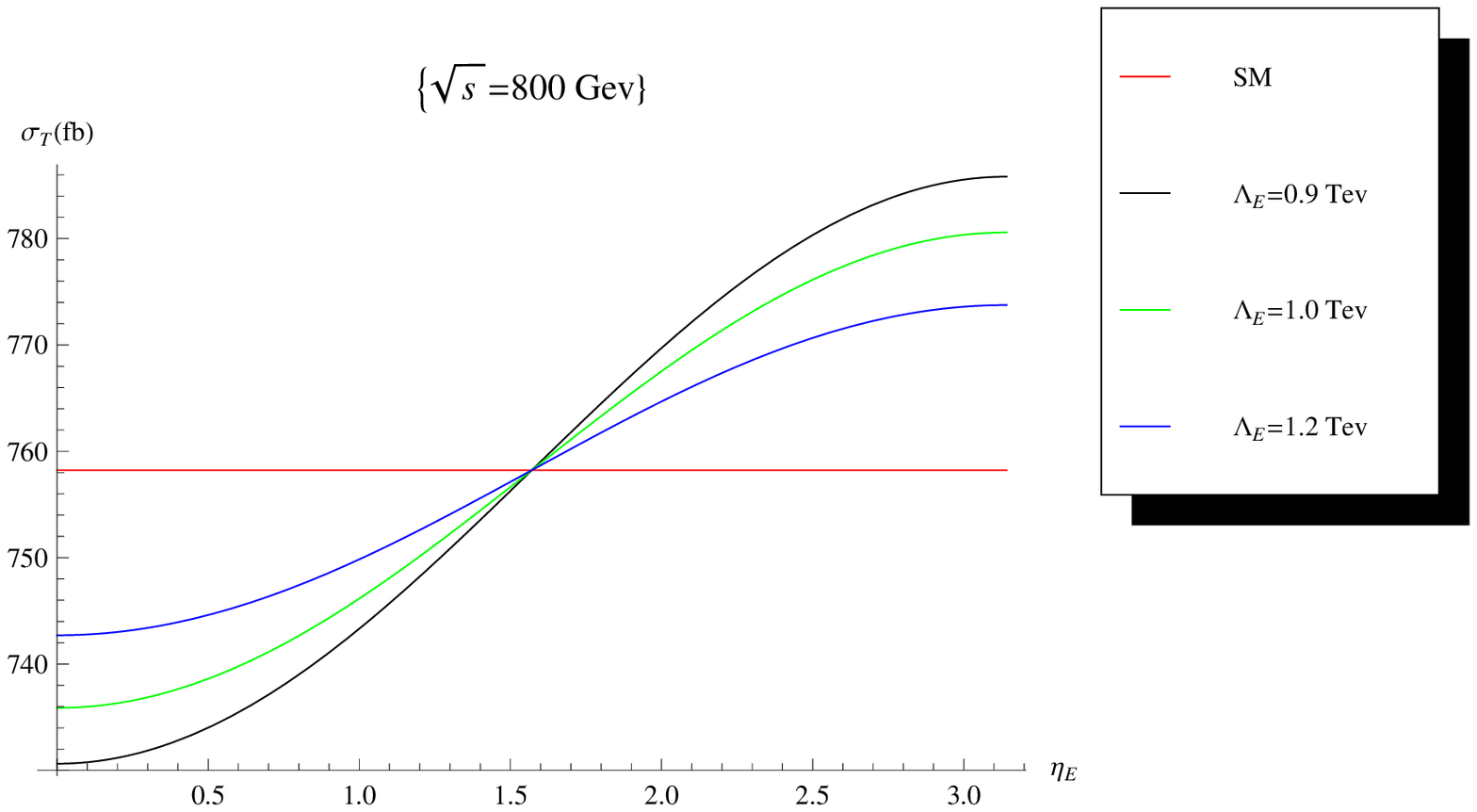}
}\vspace{-2cm}
\caption[]{Pair Annihilation:~ Time average of total cross section $(\sigma_T)$ for different
values of NC scale($\Lambda_E$).
             }
\label{fig:coordinates}
\end{minipage}
&\quad
\begin{minipage}[c]{80mm}
\centerline{
\includegraphics[width=85mm, height=95mm, angle=0]{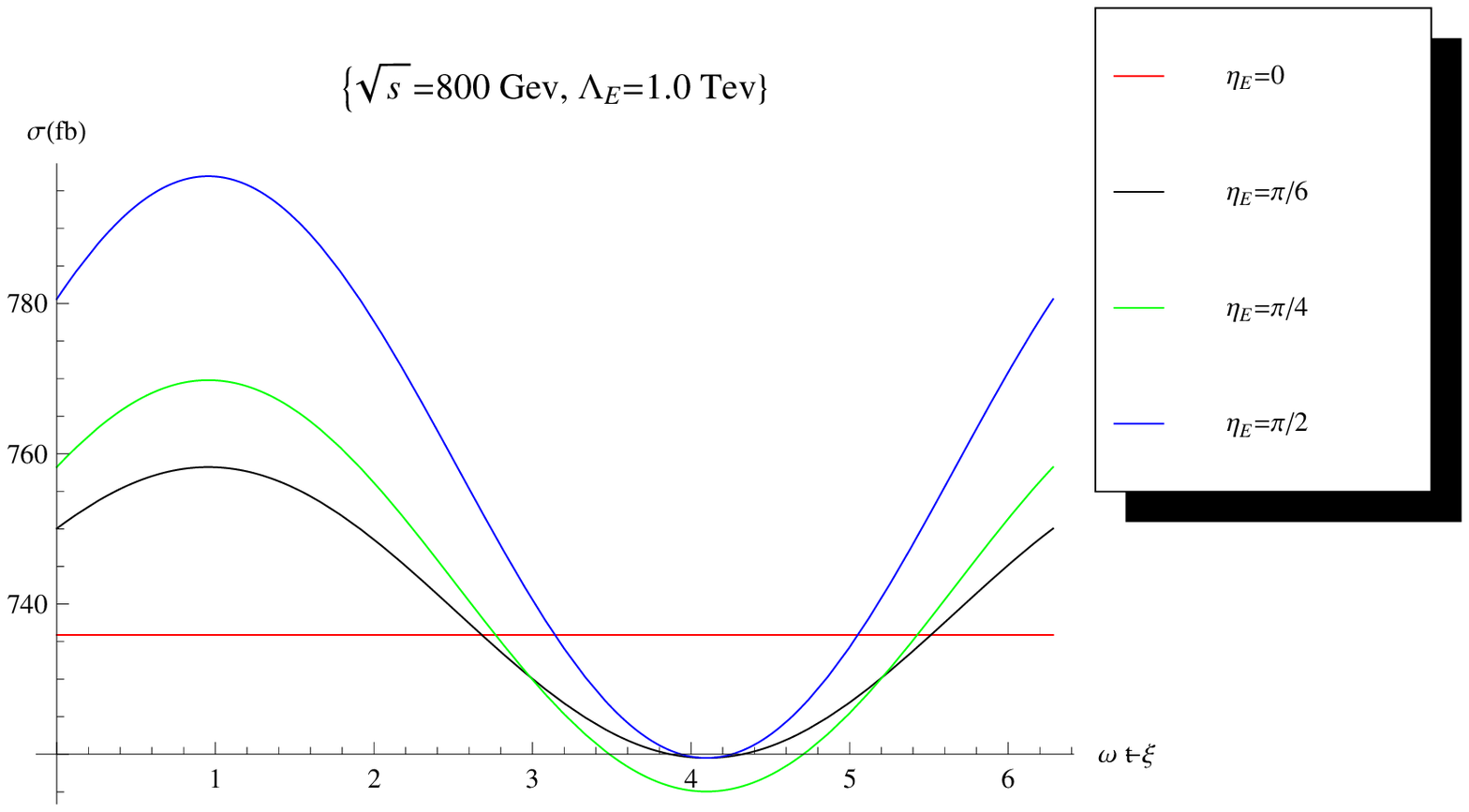}
}\vspace{-2cm}
\caption[]{Pair Annihilation:~Time dependent total cross section $(\sigma)$ vs phase($\omega t -\xi$)
for different values of angle $\eta_E$.}
\label{fig:localframe}
\end{minipage}
\end{tabular}
\end{minipage}
\end{center}
\end{figure}
\newpage


\begin{thebibliography}{99}
\bibitem{bergwitten}
  N.~Seiberg and E.~Witten,
  JHEP {\bf 9909}, 032 (1999)
  [arXiv:hep-th/9908142].

\bibitem{snyderUV}
  H.~S.~Snyder,
  Phys.\ Rev.\  {\bf 71}, 38 (1947).

\bibitem{Minwalla} S.~Minwalla, M.~Van Raamsdonk and N.~Seiberg,
  JHEP {\bf 0002}, 020 (2000)
  [arXiv:hep-th/9912072].



\bibitem{Susskind} A.~Matusis, L.~Susskind and N.~Toumbas,
  JHEP {\bf 0012}, 002 (2000)
  [arXiv:hep-th/0002075].



\bibitem{Alvarez} L.~Alvarez-Gaume and M.~A.~Vazquez-Mozo,
  Nucl.\ Phys.\  B {\bf 668}, 293 (2003)
  [arXiv:hep-th/0305093].



\bibitem{Jaeckel1} J.~Jaeckel, V.~V.~Khoze and A.~Ringwald,
  JHEP {\bf 0602}, 028 (2006)
  [arXiv:hep-ph/0508075].


\bibitem{Jaeckel2} S.~A.~Abel, J.~Jaeckel, V.~V.~Khoze and A.~Ringwald,
  JHEP {\bf 0609}, 074 (2006)
  [arXiv:hep-ph/0607188].



\bibitem{Horvat}  R.~Horvat and J.~Trampetic,
  JHEP {\bf 1101}, 112 (2011)
  [arXiv:1009.2933 [hep-ph]].


\bibitem{Wess1} B.~Jurco, S.~Schraml, P.~Schupp and J.~Wess,
  Eur.\ Phys.\ J.\  C {\bf 17}, 521 (2000)
  [arXiv:hep-th/0006246].


\bibitem{Wess2}  B.~Jurco, L.~Moller, S.~Schraml, P.~Schupp and J.~Wess,
  Eur.\ Phys.\ J.\  C {\bf 21}, 383 (2001)
  [arXiv:hep-th/0104153].



\bibitem{Calmetetal}
  X.~Calmet, B.~Jurco, P.~Schupp, J.~Wess and M.~Wohlgenannt,
  Eur.\ Phys.\ J.\  C {\bf 23}, 363 (2002)
  [arXiv:hep-ph/0111115].

\bibitem{NGetal}
  W.~Behr, N.~G.~Deshpande, G.~Duplancic, P.~Schupp, J.~Trampetic and J.~Wess,
  Eur.\ Phys.\ J.\  C {\bf 29}, 441 (2003)
  [arXiv:hep-ph/0202121].

\bibitem{Duplancicetal}
  G.~Duplancic, P.~Schupp and J.~Trampetic,
  Eur.\ Phys.\ J.\  C {\bf 32}, 141 (2003)
  [arXiv:hep-ph/0309138].


\bibitem{ggotheta}
  T.~Ohl and J.~Reuter,
  Phys.\ Rev.\  D {\bf 70}, 076007 (2004)
  [arXiv:hep-ph/0406098].

\bibitem{Melicetal1}
  B.~Melic, K.~Passek-Kumericki, J.~Trampetic, P.~Schupp and M.~Wohlgenannt,
  Eur.\ Phys.\ J.\  C {\bf 42}, 483 (2005)
  [arXiv:hep-ph/0502249].

\bibitem{Melicetal2}
  B.~Melic, K.~Passek-Kumericki, J.~Trampetic, P.~Schupp and M.~Wohlgenannt,
  Eur.\ Phys.\ J.\  C {\bf 42}, 499 (2005)
  [arXiv:hep-ph/0503064].




\bibitem{Dasetal}
  P.~K.~Das, N.~G.~Deshpande and G.~Rajasekaran,
  Phys.\ Rev.\  D {\bf 77}, 035010 (2008)
  [arXiv:0710.4608 [hep-ph]].



\bibitem{Hewett01}J.Hewett, F.J.Petriello and T.G.Rizzo, hep-ph/0201275;  J.Hewett \etal, \PRD(64,075012,2001), \PRD(66,036001,2002).





\bibitem{Schuppgneutr}
  P.~Schupp, J.~Trampetic, J.~Wess and G.~Raffelt,
   ``The photon neutrino interaction in non-commutative gauge field theory  and
  Eur.\ Phys.\ J.\  C {\bf 36}, 405 (2004)
  [arXiv:hep-ph/0212292].



\bibitem{Haghighatgneutr}
  M.~Haghighat, M.~M.~Ettefaghi and M.~Zeinali,
  Phys.\ Rev.\  D {\bf 73}, 013007 (2006)
  [arXiv:hep-ph/0511042].



\bibitem{Najafabadipoltopqur}
  M.~Mohammadi Najafabadi,
   ``Semi-Leptonic Decay of a Polarized Top Quark in the Noncommutative Standard
  Phys.\ Rev.\  D {\bf 74}, 025021 (2006)
  [arXiv:hep-ph/0606017].

\bibitem{MahajantbW}
  N.~Mahajan,
  Phys.\ Rev.\  D {\bf 68}, 095001 (2003)
  [arXiv:hep-ph/0304235].


\bibitem{IltanZdecays}
  E.~O.~Iltan,
   ``The $Z^- \to \ell^{+} \ell^{-}$ and $W \to \nu_{\ell} \ell^{+}$ decays in
  Phys.\ Rev.\  D {\bf 66}, 034011 (2002)
  [arXiv:hep-ph/0204332].

\bibitem{Deshpandetriplecoup}
  N.~G.~Deshpande and X.~G.~He,
  Phys.\ Lett.\  B {\bf 533}, 116 (2002)
  [arXiv:hep-ph/0112320].



\bibitem{Najafabaditopqur}
  M.~M.~Najafabadi,
  Phys.\ Rev.\  D {\bf 77}, 116011 (2008)
  [arXiv:0803.2340 [hep-ph]].


\bibitem{Trampeticpheno}
  J.~Trampetic,
   ``Renormalizability and Phenomenology of theta-expanded Noncommutative Gauge
  Fortsch.\ Phys.\  {\bf 56}, 521 (2008)
  [arXiv:0802.2030 [hep-ph]].

\bibitem{BuricZgg}
  M.~Buric, D.~Latas, V.~Radovanovic and J.~Trampetic,
  Phys.\ Rev.\  D {\bf 75}, 097701 (2007).





\bibitem{Abbiendipairann}
  G.~Abbiendi {\it et al.}  [OPAL Collaboration],
  Phys.\ Lett.\  B {\bf 568}, 181 (2003)
  [arXiv:hep-ex/0303035].

\bibitem{Melicquarkonia}
  B.~Melic, K.~Passek-Kumericki and J.~Trampetic,
   ``Quarkonia decays into two photons induced by the space-time
  Phys.\ Rev.\  D {\bf 72}, 054004 (2005)
  [arXiv:hep-ph/0503133].



\bibitem{Melickpi}
  B.~Melic, K.~Passek-Kumericki and J.~Trampetic,
  Phys.\ Rev.\  D {\bf 72}, 057502 (2005)
  [arXiv:hep-ph/0507231].

\bibitem{Alboteanuothsq}
  A.~Alboteanu, T.~Ohl and R.~Ruckl,
  Phys.\ Rev.\  D {\bf 76}, 105018 (2007)
  [arXiv:0707.3595 [hep-ph]].


\bibitem{Prasanta1}
  A.~Prakash, A.~Mitra and P.~K.~Das,
  Phys.\ Rev.\  D {\bf 82}, 055020 (2010)
  [arXiv:1009.3554 [hep-ph]].

\bibitem{Prasanta2}
  P.~K.~Das, A.~Prakash and A.~Mitra,
  Phys.\ Rev.\  D {\bf 83}, 056002 (2011)
  [arXiv:1009.3571 [hep-ph]].

\bibitem{wang}
  W.~Wang, F.~Tian and Z.~M.~Sheng,
  arXiv:1105.0252 [hep-ph].






\bibitem{AlboteanuLHC}
  A.~Alboteanu, T.~Ohl and R.~Ruckl,
  Phys.\ Rev.\  D {\bf 74}, 096004 (2006)
  [arXiv:hep-ph/0608155].

\bibitem{AlboteanuILC}
  A.~Alboteanu, T.~Ohl and R.~Ruckl,
  eConf {\bf C0705302}, TEV05 (2007)
  [Acta Phys.\ Polon.\  B {\bf 38}, 3647 (2007)]
  [arXiv:0709.2359 [hep-ph]].




\bibitem{earthroteffs1}
  H.~Grosse and Y.~Liao,
  Phys.\ Rev.\  D {\bf 64}, 115007 (2001)
  [arXiv:hep-ph/0105090].

\bibitem{earthroteffs2}
  Y.~Liao and C.~Dehne,
  Eur.\ Phys.\ J.\  C {\bf 29}, 125 (2003)
  [arXiv:hep-ph/0211425].


\bibitem{earthroteffs3}
  J.~i.~Kamoshita,
  Eur.\ Phys.\ J.\  C {\bf 52}, 451 (2007)
  [arXiv:hep-ph/0206223].

\bibitem{earthroteffs4}
  M.~Haghighat, N.~Okada and A.~Stern,
  Phys.\ Rev.\  D {\bf 82}, 016007 (2010)
  [arXiv:1006.1009 [hep-ph]].

\bibitem{ILC1}
  J.~(.~).~Brau {\it et al.}  [ILC Collaboration],
  arXiv:0712.1950 [physics.acc-ph].

\bibitem{ILC2}
  G.~Aarons {\it et al.}  [ILC Collaboration],
  arXiv:0709.1893 [hep-ph].

\bibitem{Bichl1} A.~Bichl, J.~Grimstrup, H.~Grosse, L.~Popp, M.~Schweda and R.~Wulkenhaar,
  JHEP {\bf 0106}, 013 (2001)
  [arXiv:hep-th/0104097].



\bibitem{Martin}  C.~P.~Martin,
  Nucl.\ Phys.\  B {\bf 652}, 72 (2003)
  [arXiv:hep-th/0211164].


\bibitem{Martin-Tam1} C.~P.~Martin and C.~Tamarit,
  Phys.\ Rev.\  D {\bf 72}, 085008 (2005)
  [arXiv:hep-th/0503139].


\bibitem{Buric1} M.~Buric, V.~Radovanovic and J.~Trampetic,
  JHEP {\bf 0703}, 030 (2007)
  [arXiv:hep-th/0609073].


\bibitem{Buric2} D.~Latas, V.~Radovanovic and J.~Trampetic,
  Phys.\ Rev.\  D {\bf 76}, 085006 (2007)
  [arXiv:hep-th/0703018].



\bibitem{Buric3}  M.~Buric, D.~Latas, V.~Radovanovic and J.~Trampetic,
  Phys.\ Rev.\  D {\bf 77}, 045031 (2008)
  [arXiv:0711.0887 [hep-th]].



\bibitem{Martin-Tam2} C.~P.~Martin and C.~Tamarit,
  Phys.\ Lett.\  B {\bf 658}, 170 (2008)
  [arXiv:0706.4052 [hep-th]].


\bibitem{Ettefaghi} M. M. Ettefaghi, M. Haghighat and R. Mohammadi, Phys.Rev. \textbf{D} 82, 105017 (2010). 

\bibitem{Buric4}  M.~Buric, D.~Latas, V.~Radovanovic and J.~Trampetic,
  Phys.\ Rev.\  D {\bf 83}, 045023 (2011)
  [arXiv:1009.4603 [hep-th]].


    
\bibitem{Horvat2} R.~Horvat, D.~Kekez and J.~Trampetic,
  Phys.\ Rev.\  D {\bf 83}, 065013 (2011)
  [arXiv:1005.3209 [hep-ph]].


\bibitem{Horvat3} R.~Horvat, D.~Kekez, P.~Schupp, J.~Trampetic and J.~You,
  arXiv:1103.3383 [hep-ph].



\bibitem{feyncalc}
  R.~Mertig, M.~Bohm and A.~Denner,
  Comput.\ Phys.\ Commun.\  {\bf 64}, 345 (1991).

\bibitem{FORM}
  J.~A.~M.~Vermaseren,
  arXiv:math-ph/0010025.



\end{thebibliography}
\end{document}